\documentclass[9pt,technote]{IEEEtran}

\def \be {\begin{equation}}
\def \ee {\end{equation}}

\def \nn {\nonumber}

\usepackage{ifpdf}

\usepackage[dvips]{graphicx}
\usepackage{color}

\usepackage[cmex10]{amsmath}
\usepackage{amssymb}

\begin{document}
%
\title{\LARGE \bf Dynamic Modularity Approach to Adaptive Inner/Outer Loop Control of Robotic Systems}
%
%
%

\author{Hanlei~Wang, Wei Ren, Chien Chern Cheah, Yongchun Xie, and  Shangke Lyu
\thanks{H. Wang and Y. Xie are with the Science and Technology on Space Intelligent Control Laboratory,
Beijing Institute of Control Engineering,
Beijing 100094, China (e-mail: hlwang.bice@gmail.com; xieyongchun@vip.sina.com).

W. Ren is with the Department of Electrical and Computer Engineering, University of California, Riverside, CA 92521 USA (e-mail: ren@ee.ucr.edu).

C. C. Cheah and S. Lyu are with the School of Electrical and Electronic Engineering, Nanyang Technological University, Singapore 639798, Republic of Singapore (e-mail: ecccheah@ntu.edu.sg; lyus0001@e.ntu.edu.sg).
}
}
\maketitle

\begin{abstract}
Modern applications of robotics typically involve a robot control system with an inner PI (proportional-integral) or PID (proportional-integral-derivative) control loop and an outer user-specified control loop. The existing outer loop controllers, however, do not take into consideration the dynamic effects of robots and their effectiveness relies on the ad hoc assumption that the inner PI or PID control loop is fast enough, and other torque-based control algorithms cannot be implemented in robotics with closed architecture. This paper investigates the adaptive control of robotic systems with an inner/outer loop structure, taking into full account the effects of the dynamics and the system uncertainties, and both the task-space control and joint-space control are considered. We propose a dynamic modularity approach to resolve this issue, and a class of adaptive outer loop control schemes is proposed and their role is to dynamically generate the joint velocity (or position) command for the low-level joint servoing loop. Without relying on the ad hoc assumption that the joint servoing is fast enough or the modification of the low-level joint controller structure, we rigorously show that the proposed outer loop controllers can ensure the stability and convergence of the closed-loop system. We also propose the outer loop versions of several standard joint-space direct/composite adaptive controllers for rigid or flexible-joint robots, and a promising conclusion may be that most torque-based adaptive controllers for robots can be designed to fit the inner/outer loop structure by using the new definition of the joint velocity (or position) command. Simulation results are provided to show the performance of various adaptive outer loop controllers, using a three-DOF (degree-of-freedom) manipulator, and experiment results using the UR10 robotic system are also presented.

\end{abstract}


\section{Introduction}

User-friendliness is an important aspect of modern automatic machines, especially if they are expected to do extensive work in cooperation with human beings. The control systems for modern robotic systems, unfortunately, have not yet reach this expectation though numerous control algorithms have been developed over the past several decades. To serve this purpose, the controlled robotic system might have to be reliable, robust, and flexible to satisfy the user's needs. As we take a deep look at the development of computers (for instance, the recent hybrid computers---Surface Pro 4, iPad Pro, etc.), the module design plays a vital role in promoting their success in our everyday life. The relationship between the operating systems (e.g., the Windows System) and the application programs further validates the desirability of this module design philosophy.

Historically, the applications of robotics have undergone the following phases:
\begin{enumerate}

\item Traditional industrial applications---factory automation in a structured environment with a simple joint-space PID (proportional-integral-derivative) inner loop control as well as inverse kinematics;

\item Modern applications---beyond factory automation in an unstructured environment with sensory feedback in the task space, e.g., (outer loop) visual servoing and task-space control, which, however, either cannot be implemented in robotics with closed architecture or rely on the ad hoc assumption that the combination of the inner and outer loops is stable and the effect of the dynamics can be neglected. This ad hoc assumption typically (approximately) holds only on the occasion that the given task is slow enough.

\end{enumerate}
The inner/outer loop structure of robotic systems (e.g., most commercial robotic systems) enjoys certain module design flavor and has some desirable properties, e.g., it is beneficial for generating a high joint stiffness by employing a fast inner joint servoing while it is not reliable as directly specifying the control torque due to the limitation of the communication (generally required for exerting a coupling control action) reliability. For this reason, the gap between the study of advanced robot control theory and practical applications is longstanding. In the academic field, most advanced controller designs are torque-based and typically require an open torque control loop. In practical applications, the much more reliable and robust velocity control mode is adopted. It seems necessary here to recall the {standard inner/outer loop structure} of most commercial (industrial) robotic systems: 1) an outer (kinematic) loop using a centralized computer with enough computing power;  2) an inner dynamic loop that consists of $n$ independent processors equipped at each joint in a decentralized way (i.e., each processor only receives/sends signals from/to its local sensors, actuator, and the centralized computer). The inner dynamic loop usually evolves at a much faster sampling than the outer (kinematic) loop.

The attempts that aim to address the control of this kind of robotic systems in the task space occur in, e.g., \cite{Aicardi1995_ICRA,Roy2002_TRA,Kelly2005_AUT,Camarillo2008_Robotica}. However, these controllers cannot ensure the tracking error convergence without significantly modifying the low-level PI (proportional-integral) controller to be a more complex one. One may note that these control algorithms are all based on the standard resolved motion rate control proposed by \cite{Whitney1969_Resolved} to design the joint velocity command. Then what actually prevents the application of advanced robot controllers? Let us first retrospect the realization of the standard computed torque controller that has been discussed in \cite[p.~209, p.~210]{Spong1989_Book} (see also the trajectory precorrection based on the computed torque feedforward in \cite{Grotjahn2002_IJRR}). The specific procedure in \cite{Spong1989_Book}  is to modify the terms in the feedforward action that may involve coupling by replacing certain signals (positions or velocities) with their desired values. In this way, no communication between the joint processors at the dynamic servoing loop is required since the desired values of all the joints are stored in each joint's computer a priori (which would, on the other hand, mean that this algorithm is not flexible in the case that the desired trajectory is subjected to changes). The overall impression is that for the sake of reducing computational burden, this scheme performs the feedforward at the joint control loop in a relatively limited manner and with many nonlinear terms being neglected. Another important well-recognized reason is that most commercial/industrial robotic systems do not have an open torque control loop (see, e.g., \cite{Sanfilippo2015_RAM}). These two factors give rise to the awkward situation of the modern torque-based robot control algorithms (e.g., the adaptive algorithms in \cite{Slotine1987_IJRR,Spong1989_SCL,Cheah2006_TAC,Liu2006_TRO,Dixon2007_TAC,Wang2015_AUT,Wang2012a_MECH} and the robust algorithms in \cite{Spong1992_TAC,Yao1996_ASME}), i.e., it is hard to apply these algorithms to robotic systems with an inner/outer loop structure. The precorrection scheme in \cite{Grotjahn2002_IJRR}, by producing a trajectory correction term based on the
inverse manipulator dynamics and then adding it to the desired joint trajectory, improves the performance of the industrial robotic system without modifying the low-level controller structure. The main proofs of the performance improvement, however, are by the experimental results rather than by the rigorous analysis.

In this paper, we rigorously address this issue in the context of adaptive task-space/joint-space control for robotic systems with an embedded low-level PI joint velocity controller (or PID joint position controller) and with uncertain dynamics (and kinematics). The use of PI velocity controller or PID position controller in most industrial/commercial robotic systems is well recognized (see, e.g., \cite{Caccavale1994_CEP,Swevers2007_CSM}). {Our main purpose here is to develop a class of adaptive outer loop controllers that can ensure the stability and convergence of the robotic systems with the dynamic effect being taken into full account and without modifying the embedded inner PI or PID control loop}. The application of the current adaptive (or robust) task-space regulation/tracking algorithms (e.g., \cite{Cheah2003_TRA,Cheah2006_IJRR,Liu2006_TRO,Dixon2007_TAC,Wang2012a_MECH,Liang2014_ASJC,Wang2012_TAES,Wang2015_AUT,Wang2010_TCST,Leite2011_IFAC}) to robotic systems with an inner/outer loop structure, for a long period, relies on the ad hoc assumption that the inner joint servoing loop is fast enough or the modification of the inner joint controller structure. A preliminary version of the paper was presented in \cite{Wang2016_CCC} where the simultaneous opening of the position and velocity commands is required. We here extend this preliminary result to address the case that only the position (or velocity) command is designable (which is considered to be much more common in most industrial/commercial robots), and to additionally consider the case of composite adaptation and joint flexibility as well as include the experimental results.

We first propose two adaptive task-space regulation controllers that rule out the fundamental limitations of the existing results, by dynamically incorporating an adaptively scaled dynamic compensation that exploits the physically independent nature of the low-level controller structure. The first controller, by introducing an adaptive filter, avoids the task-space velocity measurement, and the second one avoids the use of the task-space velocity by using an observer [motivated by the one in \cite{Wang2015_CCC} with a modified feedback gain (which depends on the estimated Jacobian matrix) to achieve feedback separation].
 Both of the adaptive controllers are qualified outer loop control schemes that can be applied to robotic systems with an unmodifiable joint servoing controller (PI velocity or PID position controller) (e.g., most industrial/commercial robots), taking into account the dynamic loop of the robotic systems. From a robot control perspective, most existing kinematic algorithms are not mathematically rigorous in that either the effects of the inner joint control loop are not considered (e.g., \cite{Weiss1987_JRA,Siciliano1990_Robotica,Hutchinson1996_TRA}), or the low-level joint servoing controller is assumed to be strong enough to ensure (yet cannot rigorously guarantee due to the absence of dynamic compensation action in the low-level dynamic loop) the square-integrability and boundedess of the velocity tracking error (e.g., \cite{Wang2016_TAC,Wang2015_TACsubmitted}), or even the modification of the low-level controller structure (e.g., \cite{Aicardi1995_ICRA,Roy2002_TRA,Kelly2005_AUT}). These limitations are mainly caused by the inner/outer loop structure and closed controller architecture. The proposed outer loop controller here, by dynamically
 incorporating adaptively scaled dynamic compensation action and adaptive transpose Jacobian feedback, ensures the singularity-robust stability and convergence of the task-space position error without relying on any modification of the low-level PI/PID controller structure. Due to the independence of the design of the outer loop controller and that of the low-level PI/PID controller and the injection of the dynamic compensation, the proposed design approach is referred to as \emph{dynamic modularity approach}.

 We then show that the observer-based task-space regulation scheme can be extended to the case of task-space tracking. The obtained control scheme has an interesting feature that the inverse of the estimated Jacobian matrix is used for introducing feedforward and its transpose is used for introducing feedback, in contrast to most existing task-space algorithms that only employ the inverse of the Jacobian matrix to exert feedforward and feedback actions (see, e.g., \cite{Slotine1987_IJRR,Niemeyer1991_IJRR,Cheah2006_IJRR,Wang2015_AUT}), and the benefit of this lies in two folds: 1) it yields the feedback separation of the kinematic and dynamic loops, thus reducing the activity of the dynamic compensation action; 2) it is reducible in the sense that once the desired task-space velocity becomes zero, the tracking control law reduces to the regulation control law without involving the inverse of the estimated Jacobian matrix.

Finally, we illustrate how the adaptively scaled dynamic compensation enables several typical adaptive robot controllers for {joint-space trajectory tracking} to be applicable to robotic systems with an inner/outer loop structure (e.g., most industrial/commercial robotic systems). In particular, the proposed dynamic modularity approach is further shaped to incorporate the composite adaptation for improving performance and to address the issue of joint flexibility. By these additional examples, it seems hopeful that most adaptive dynamic controllers for robots in the literature with the use of adaptively scaled dynamic compensation and new definition of the joint velocity (or position) command would be rendered to be qualified outer loop schemes. {Another potential favorable point may be the reduction of the cost of the laboratory experimental research on advanced robot control theory in that there no longer needs to develop a specific manipulator with an open torque control loop (usually requiring extensive efforts and time) and any commercial robot (cost-efficient due to the large-scale production) can be directly used}.

\section{Preliminaries}

\subsection{Background and Motivation}

Inner/outer loop control is typical in modern applications of robotic systems (see Fig. 1), and generally the inner loop is designed by the robot production company and closed and only the outer loop is open to the user. The user can specify the position or velocity command within the outer loop based on measurements in joint space [and task space (e.g., image space)], which is then sent to the inner loop as a reference signal. The main benefit of adopting an inner/outer loop structure may be that the inner loop (due to simplicity) can be operated at a high sampling rate and thus a high stiffness can be maintained, and that the outer loop can be operated at a low sampling rate allowing relatively complicated communication and sensing. In addition, inner/outer loop structure does help to realize the relative independence of the manipulator production company and the users, and to promote the large-scale production of manipulators due to the invariance of the inner control loop.
\begin{figure}
\centering
\begin{minipage}[t]{1.0\linewidth}
\centering
\includegraphics[width=3.2in]{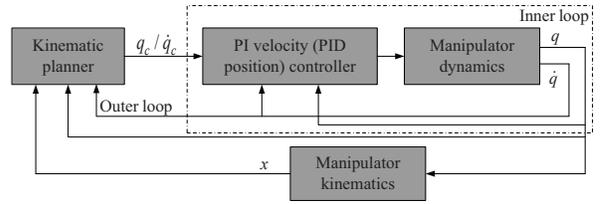}
\caption{Inner/outer loop control ($q_c$ is the position command, $\dot q_c$ is the velocity command, and $q$, $\dot q$, and $x$ are the joint position, joint velocity, and task-space position, respectively).}\label{fig:side:a}
\end{minipage}%
\end{figure}

Historically, most theoretical results on adaptive manipulator control are presented in the context that the joint torque is directly designable rather than in the framework of the inner/outer loop. This results in the longstanding gap between the study of advanced robot control theory and the applications of robots. The existing results (e.g., \cite{Aicardi1995_ICRA,Roy2002_TRA,Kelly2005_AUT,Grotjahn2002_IJRR,Weiss1987_JRA,Siciliano1990_Robotica,Hutchinson1996_TRA}) are either ad hoc, e.g., effective under the assumption that the joint servoing is fast enough, relying on the modification of the low-level controller structure which is unmodifiable in practice, or not theoretically rigorous.

{Our main purpose is to develop a class of adaptive outer loop controllers using the task-space and joint-space sensory measurement (instead of obtaining the task-space information based on the kinematics as in Fig. 1 since the kinematics is unknown) to take place of the kinematic planner in Fig. 1 so that the stability and convergence of the closed-loop robotic systems with uncertain dynamics (and kinematics) can rigorously be ensured, and to finally approach the goal of ``modularity''}.

\subsection{Manipulator Kinematics and Dynamics}

Consider an $n$-DOF (degree-of-freedom) manipulator actuated by permanent magnet DC motors. Let $x\in R^m$ be the position of the end-effector in the task space and it is relevant to the joint position by the following nonlinear mapping \cite{Craig2005_Book,Spong2006_Book}
\be
\label{eq:1}
x=f(q)
\ee
where $q\in R^n$ denotes the joint position and $f: R^n\to R^m$ is the mapping from joint space to task space. We here assume that $n\ge m$, i.e., the manipulator can either be nonredundant or redundant.

Differentiating (\ref{eq:1}) with respect to time yields the relation between the task-space velocity $\dot x$ and joint velocity $\dot q$ \cite{Craig2005_Book,Spong2006_Book}
\be
\label{eq:2}
\dot x=J(q)\dot q
\ee
where $J(q)\in R^{m\times n}$ is the Jacobian matrix. If the kinematic parameters are unknown, the task-space position/velocity can no longer be derived by the direct kinematics given above. The typical practice in this case is to employ certain task-space sensors (e.g., a camera) to obtain the task-space position information.

The dynamics of the manipulator can be written as \cite{Spong2006_Book}
\be
\label{eq:3}
M(q)\ddot q+C(q,\dot q)\dot q+B \dot q+g(q)=K u
\ee
where $M(q)\in R^{n\times n}$ is the inertia matrix, $C(q,\dot q)\in R^{n\times n}$ is the Coriolis and centrifugal matrix, $B\in R^{n\times n}$ is a constant diagonal positive definite matrix, $g(q)\in R^n$ is the gravitational torque, $u\in R^n$ is the armature voltage, and $K\in R^{n\times n}$ is a constant diagonal positive definite matrix.

Four basic properties associated with (\ref{eq:2}) and (\ref{eq:3}) that shall be useful for the controller design and stability analysis are listed as follows.

\emph{Property 1 (\cite{Cheah2006_IJRR}):} The kinematics (\ref{eq:2}) depends linearly on a constant kinematic parameter vector $a_k$, which gives rise to
\be
\label{eq:4}
J(q)\psi=Y_k(q,\psi)a_k
\ee
where $\psi\in R^n$ is a vector and $Y_k(q,\psi)$ is the kinematic regressor matrix.

\emph{Property 2 (\cite{Slotine1991_Book,Spong2006_Book}):} The inertia matrix $M(q)$ is symmetric and uniformly positive definite.

\emph{Property 3 (\cite{Slotine1991_Book,Spong2006_Book}):} The matrix $C(q,\dot q)$ can be appropriately defined such that the matrix $\dot M(q)-2C(q,\dot q)$ is skew-symmetric.

\emph{Property 4 (\cite{Slotine1991_Book,Spong2006_Book}):} The dynamics (\ref{eq:3}) depends linearly on a constant dynamic parameter vector $a_d$, which yields
\be
\label{eq:5}
M(q)\dot \zeta+C(q,\dot q)\zeta+B\zeta+g(q)=Y_d(q,\dot{q},\zeta,\dot \zeta)a_d
\ee
where $\zeta\in R^n$ is a differentiable vector, $\dot\zeta$ is the time derivative of $\zeta$, and $Y_d(q,\dot{q},\zeta,\dot \zeta)$ is the dynamic regressor matrix.

\section{Adaptive Inner/Outer Loop Control}

Kinematic control typically appears in the context of inner/outer loop control of robot manipulators and its focus is on the design of the joint velocity (or position) command. Historically, the effectiveness of kinematic control stands on the relatively strong assumption that the inner control (PI velocity control or PID position control) loop is fast enough so that the dynamic effects of the inner loop can be neglected. Here, we present a dynamic modularity approach to ensure the convergence of the task-space position error, without relying on this ad hoc assumption or the modification of the inner control loop.

Let $x_d\in R^m$ denote the desired task-space position. For the regulation problem, it is set as constant; for the tracking problem, it can be set as time-varying and in this case, we assume that $x_d$, $\dot x_d$, and $\ddot x_d$ are all bounded.

\subsection{Filter-Based Adaptive Regulation Control}

Consider the case that the inner control loop employs a PI velocity control, and the PI gains of the inner PI velocity control loop (the case of PID position control is discussed in Sec. III-D) are supposed to be $K_P$ and $K_I$ (diagonal and positive definite). Then, the PI action can be written as
\be
\label{eq:6}
u=-K_P (\dot q-\dot q_c)-K_I \left(q-q_c\right).
\ee

To avoid the task-space velocity measurement, we introduce the following adaptive passive filter
\be
\label{eq:7}
\dot y=-K_1 y+ K_1 \hat J^{T}(q)\Delta x,
\ee
upon which, we define the joint reference velocity as
\be
\label{eq:8}
\dot q_r=-K_2 y
\ee
where $\Delta x=x-x_d$, $K_1$ and $K_2$ are diagonal positive definite matrices, and $\hat J(q)$ is the estimate of the Jacobian matrix $J(q)$ and is obtained by replacing $a_k$ in $J(q)$ with its estimate $\hat a_k$.
Define a sliding vector
\be
\label{eq:9}
s=\dot q-\dot q_r.
\ee
 Substituting (\ref{eq:8}) into (\ref{eq:2}) and using Property 1 gives
\begin{align}
\label{eq:10}
\dot x=&-\hat J(q)K_2 y-[\hat J(q)-J(q)]\dot q-\hat J(q)\dot q_r+\hat J(q) \dot q\nn\\
=&-\hat J(q) K_2 y-Y_k(q,\dot q)\Delta a_k+\hat J(q) s.
\end{align}
where $\Delta a_k=\hat a_k-a_k$.

We define the joint velocity command by the following dynamic system
 \begin{align}
 \label{eq:11}
 \dot q_c+\hat{\mathcal K}_I q_c=&\dot q_r+\hat{\mathcal K}_I q_r+{\rm diag}[\hat w_i,i=1,\dots,n]\nn\\
  &\times\left[-\alpha \hat J^{T}(q)\Delta x+Y_d(q,\dot q,\dot q_r,\ddot q_r) \hat a_d\right]
  \end{align}
  with
  \begin{align}
 \label{eq:12}
 q_r=&q_r(0)+\int_0^t \dot q_r(\sigma)d\sigma
 \end{align}
 where $q_r(0)$ can be chosen as an arbitrary constant vector, $\alpha$ is a positive design constant, $\hat a_d$ is the estimate of $a_d$, $\hat w_i$  denotes the scale weight, $i=1,\dots,n$, ${\rm diag}[\hat w_i,i=1,\dots,n] [-\alpha \hat J^{T}(q)\Delta x+Y_d(q,\dot q,\dot q_r,\ddot q_r) \hat a_d]$ denotes the adaptively scaled dynamic compensation action, and $\hat{\mathcal K}_I$ denotes the estimate of ${\mathcal K}_I=K_P^{-1}K_I$, which can be expressed as
\be
\label{eq:13}
\hat{\mathcal K}_I={\rm diag}[\hat w_I]
\ee
with $\hat w_I$ being an $n$-dimensional vector. The adaptation laws for $\hat a_k$, $\hat w=[\hat w_1,\dots,\hat w_n]^T$, $\hat a_d$, and $\hat w_I$ are given as
\begin{align}
\label{eq:14}
&\dot{\hat a}_k=\Gamma_k Y_k^{ T}(q,\dot q)\Delta x\\
\label{eq:15}
&\dot{\hat w}=-\Lambda {\rm diag}[-\alpha \hat J^T(q)\Delta x+Y_d(q,\dot q,\dot q_r,\ddot q_r)\hat a_d]s
\\
\label{eq:16}
&\dot{\hat a}_d=-\Gamma_d Y_d^T(q,\dot q,\dot q_r,\ddot q_r) s\\
\label{eq:17}
&\dot{\hat w}_I=\Lambda_I{\rm diag}[q_c-q_r]s
\end{align}
where $\Lambda$ and $\Lambda_I$ are diagonal positive definite matrices, and $\Gamma_k$ and $\Gamma_d$ are symmetric positive definite matrices.

\emph{Theorem 1:} Suppose that $\hat{\mathcal K}_I$ is uniformly positive definite and $\hat J(q)$ has full row rank. Then, the adaptive outer loop controller given by (\ref{eq:11}), (\ref{eq:12}), (\ref{eq:14}), (\ref{eq:15}), (\ref{eq:16}), and (\ref{eq:17}) for the robotic system (\ref{eq:2}) and (\ref{eq:3}) under the inner PI controller (\ref{eq:6}) ensures the stability of the system and convergence of the task-space position error, i.e., $\Delta x\to 0$ as $t\to\infty$.

\emph{Proof:} Substituting (\ref{eq:6}) and (\ref{eq:11}) into the manipulator dynamics (\ref{eq:3}) and taking into account Property 4 gives
\begin{align}
\label{eq:18}
M&(q)\dot s+C(q,\dot q)s\nn\\
=&-Bs-K^\ast(\dot q+{\mathcal K}_I q)+K^\ast\left(\dot q_c+\hat {\mathcal K}_Iq_c\right)\nn\\
&-K^\ast\Delta {\mathcal K}_I q_c-Y_d(q,\dot q,\dot q_r,\ddot q_r) a_d\nn\\
=&-Bs-K^\ast(\dot q+{\mathcal K}_I q)+K^\ast\left(\dot q_r+\hat{\mathcal K}_I q_r\right)\nn\\
&+K^\ast{\rm diag}[\hat w_i,i=1,\dots,n]\left[-\alpha \hat J^{T}(q)\Delta x+Y_d(q,\dot q,\dot q_r,\ddot q_r) \hat a_d\right]\nn\\
&-K^\ast\Delta {\mathcal K}_I q_c-Y_d(q,\dot q,\dot q_r,\ddot q_r) a_d\nn\\
=&-Bs-K^\ast(\dot q+{\mathcal K}_I q)+K^\ast\left(\dot q_r+{\mathcal K}_I q_r\right)\nn\\
&+K^\ast{\rm diag}[\hat w_i,i=1,\dots,n]\left[-\alpha \hat J^{T}(q)\Delta x+Y_d(q,\dot q,\dot q_r,\ddot q_r) \hat a_d\right]\nn\\
&-K^\ast\Delta {\mathcal K}_I (q_c-q_r)-Y_d(q,\dot q,\dot q_r,\ddot q_r) a_d\nn\\
=&-(K^\ast+B) s -KK_I\left[q-q(0)-\int_0^t \dot q_r(\sigma)d\sigma+\delta_0\right]\nn\\
&+K^\ast {\rm diag}[\hat w_i,i=1,\dots,n] [-\alpha \hat J^{T}(q)\Delta x+Y_d(q,\dot q,\dot q_r,\ddot q_r)\hat a_d]\nn\\
&-Y_d(q,\dot q,\dot q_r,\ddot q_r) a_d\nn\\
=&-(K^\ast+B) s-\alpha \hat J^{T}(q)\Delta x-KK_I \left[\int_0^t s(\sigma)d\sigma+\delta_0\right]\nn\\
&+K^\ast ({\rm diag}[\hat w_i-k_i^{\ast -1},i=1,\dots,n]) \nn\\
&\times[-\alpha \hat J^T(q)\Delta x+Y_d(q,\dot q,\dot q_r,\ddot q_r)\hat a_d]\nn\\
&-K^\ast\Delta {\mathcal K}_I (q_c-q_r)+Y_d(q,\dot q,\dot q_r,\ddot q_r)\Delta a_d\nn\\
=&-(K^\ast+B) s-\alpha \hat J^{ T}(q)\Delta x-KK_I \left[\int_0^t s(\sigma)d\sigma+\delta_0\right]\nn\\
&+{\rm diag}[-\alpha\hat J^T(q)\Delta x+Y_d(q,\dot q,\dot q_r,\ddot q_r)\hat a_d]K^\ast \Delta w \nn\\
&-K^\ast{\rm diag}[q_c-q_r]\Delta w_I +Y_d(q,\dot q,\dot q_r,\ddot q_r)\Delta a_d
\end{align}
where $\delta_0=q(0)-q_r(0)$ is a constant vector, $K^\ast=KK_P={\rm diag}[k_{ii}^\ast,i=1,\dots,n]$ with $k_{ii}^\ast$, $i=1,\dots,n$ being positive constants, $\Delta a_d=\hat a_d-a_d$, $\Delta w=[\hat w_1-k_{11}^{\ast-1},\dots,\hat w_n-k_{nn}^{\ast-1}]^T$, and $\Delta w_I=\hat w_I-w_I$ with the entries of $w_I$ being from the diagonal entries of ${\mathcal K}_I$. Consider the following Lyapunov function candidate
\begin{align}
V=&\alpha\Big(\frac{1}{2}\Delta x^T \Delta x+\frac{1}{2}y^TK_2 K_1^{-1} y+\frac{1}{2}\Delta a_k^T \Gamma_k^{-1}\Delta a_k\Big)\nn\\
&+\frac{1}{2}s^T M(q)s+\frac{1}{2}\left[\int_0^t s(\sigma)d\sigma+\delta_0\right]^T K K_I\left[\int_0^t s(\sigma)d\sigma+\delta_0\right]\nn\\
&+\frac{1}{2}\Delta w^T \Lambda^{-1}K^\ast \Delta w+\frac{1}{2}\Delta a_d^T \Gamma_d^{-1}\Delta a_d+\frac{1}{2}\Delta w_I^T \Lambda_I^{-1}K^\ast \Delta w_I
\end{align}
 whose derivative with respect to time along the trajectories of (\ref{eq:18}), (\ref{eq:7}), (\ref{eq:10}), (\ref{eq:14}), (\ref{eq:15}), (\ref{eq:16}), and (\ref{eq:17}) can be written as (using Property 3)
\be
\dot V=-\alpha y^T K_2 y-s^T (K^\ast +B) s\le 0.
\ee
Then we obtain that $y\in{\mathcal L}_2\cap {\mathcal L}_\infty$, $s\in{\mathcal L}_2\cap{\mathcal L}_\infty$, $\int_0^t s(\sigma)d\sigma\in{\mathcal L}_\infty$ $\Delta x\in{\mathcal L}_\infty$, $\hat w\in{\mathcal L}_\infty$, $\hat a_k\in{\mathcal L}_\infty$, $\hat a_d\in{\mathcal L}_\infty$, and $\hat w_I\in{\mathcal L}_\infty$.
From (\ref{eq:7}), we obtain that $\dot y\in{\mathcal L}_\infty$ and thus $y$ is uniformly continuous. From the properties of square-integrable and uniformly continuous functions \cite[p.~232]{Desoer1975_Book}, we obtain that $y\to0$ as $t\to\infty$. From (\ref{eq:8}), we know that $\dot q_r\in{\mathcal L}_\infty$ and thus $\dot q\in{\mathcal L}_\infty$. From (\ref{eq:14}), we obtain that $\dot{\hat a}_k\in{\mathcal L}_\infty$, giving rise to the boundedness of $\dot{\hat J}(q)$. From (\ref{eq:2}), we obtain that $\dot x\in{\mathcal L}_\infty$, and we then obtain that $\ddot y\in{\mathcal L}_\infty$ based on (\ref{eq:7}). This means that $\dot y$ is uniformly continuous and thus $\dot y \to 0 $ as $t\to\infty$ according to Barbalat's Lemma \cite{Slotine1991_Book}. From  (\ref{eq:7}), we then obtain that $\hat J^{T}(q)\Delta x\to 0$ as $t\to\infty$.  This means that $\Delta x\to 0$ as $t\to\infty$ since $\hat J(q)$ has full row rank. Furthermore, the result that $\dot y\in{\mathcal L}_\infty$ yields the conclusion that $\ddot q_r\in{\mathcal L}_\infty$. Rewrite (\ref{eq:11}) as
\begin{align*}
 &(\dot q_c-\dot q_r)+\hat{\mathcal K}_I (q_c-q_r)\nn\\
 &={\rm diag}[\hat w_i,i=1,\dots,n]\left[-\alpha \hat J^{T}(q)\Delta x+Y_d(q,\dot q,\dot q_r,\ddot q_r) \hat a_d\right],
  \end{align*}
  and it can be directly shown that $q_c-q_r\in{\mathcal L}_\infty$ and $\dot q_c-\dot q_r\in{\mathcal L}_\infty$ since $\hat {\mathcal K}_I$ is uniformly positive definite. We then obtain that $\dot q_c\in{\mathcal L}_\infty$.
From (\ref{eq:18}), we obtain that $\dot s\in{\mathcal L}_\infty$ using Property 2, and further that $\ddot q\in{\mathcal L}_\infty$. Hence $s$ is uniformly continuous, yielding the result that $s\to 0$ as $t\to\infty$ according to the properties of square-integrable and uniformly continuous functions \cite[p.~232]{Desoer1975_Book}. This immediately gives the conclusion that $\dot q\to 0$ as $t\to\infty$ since $\dot q_r=-K_2 y\to 0$ as $t\to\infty$. \hfill {\small $\blacksquare$}

\emph{Remark 1:} The introduction of the adaptive scales $\hat w$ and $\hat w_I$ is to accommodate the uncertain diagonal matrices $K^\ast=KK_P$ and $K_P^{-1}K_I$, and their uncertainty comes from both the actuator model and low-level controller design (generally performed by the robot production company). The part due to the actuator model (i.e., $K$) is inherently uncertain and may possibly be subjected to slow variation. The uncertainty of the part due to the low-level controller design (i.e., $K_P$ and $K_I$) is a business strategy for protecting the intellectual property right and thus it is and will be impossible to be disclosed thoroughly in the short run. In addition, the adaptive scale $\hat w$ used here is computationally efficient and free of computational singularity since it does not involve the computation of inverse of an estimated quantity (which, however, would be encountered if we directly estimate the matrix $K^\ast$). Similar techniques for handling the uncertainty of the diagonal torque-constant matrix (which describes the relation between the torque and current) appear in the context of adaptive control for rigid-link electrically-driven robots or robots with actuator uncertainty, yet with an open controller structure (i.e., the voltage can be directly specified by the user) \cite{Bridges1993_CDC,Cheah2006_TAC}.

\emph{Remark 2:} The uniform positive definiteness of $\hat {\mathcal K}_P$ can be conveniently ensured by using the projection algorithms \cite{Ioannou1996_Book}, and the full row rank of $\hat J(q)$ can be ensured by the assumption of being away from the singular configuration and the use of the projection algorithms \cite{Cheah2006_IJRR,Dixon2007_TAC}.

\emph{Remark 3: } The adaptive filter (\ref{eq:7}) with $K_1\hat J^{ T}(q)\Delta x $ as the input can be considered as an extension of \cite{Berghuis1993_SCL} to address the avoidance of task-space velocity measurement in the context of adaptive task-space control. Clearly, the joint velocity command given by (\ref{eq:11}) no longer involves the task-space velocity measurement since $\ddot q_r=-K_2\dot y$ and $\dot y$ given by (\ref{eq:7}) does not involve the task-space velocity.

\subsection{Observer-Based Adaptive Regulation Control}

The task-space observer is given as
\begin{align}
\label{eq:21}
\dot x_o=\hat J(q)\dot q_r-\beta\hat J(q)\hat J^{T}(q)\Delta x_o
\end{align}
where $x_o\in R^{m}$ denotes the observed quantity of $x$, $\Delta x_o=x_o-x$, and $\beta$ is a positive design constant, and the joint reference velocity $\dot q_r$ is now defined as
\be
\label{eq:22}
\dot q_r=-\hat J^{T}(q)\left[\gamma(x_o-x_d)\right]
\ee
where $\gamma$ is a positive design constant. {The observer (\ref{eq:21}) is motivated by \cite{Wang2015_CCC} yet with a new feedback gain $\beta\hat J(q)\hat J^{T}(q)$ for the purpose of achieving feedback separation, and the idea behind is that the actual joint velocity finally approaches the joint reference velocity $\dot q_r$ and thus $\hat J(q)\dot q_r$ would approach the estimated task-space velocity $\hat J(q)\dot q$. The desirable point is that the observer no longer depends on the joint velocity and is thus not influenced by the noise of the velocity signal.} The joint velocity command is defined as
 \begin{align}
 \label{eq:23}
 \dot q_c+\hat {\mathcal K}_I q_c=&\dot q_r+\hat {\mathcal K}_I q_r+{\rm diag}[\hat w_i,i=1,\dots,n]\nn\\
  &\times Y_d(q,\dot q,\dot q_r,\ddot q_r) \hat a_d
 \end{align}
 which no longer needs the scaled dynamic compensation action ${\rm diag}[\hat w_i,i=1,\dots,n][-\alpha \hat J^{T}(q) \Delta x]$. The adaptation laws for $\hat a_k$ and $\hat w$ are given as
 \begin{align}
 \label{eq:24}
 &\dot {\hat a}_k=\Gamma_kY_k^{T}(q,\dot q)(\Delta x-\Delta x_o)\\
 \label{eq:25}
 &\dot {\hat w}=-\Lambda{\rm diag}[Y_d(q,\dot q,\dot q_r,\ddot q_r)\hat a_d]s
 \end{align}
 The adaptation laws for $\hat a_d$ and $\hat w_I$ remain the same as (\ref{eq:16}) and (\ref{eq:17}), respectively.

Combining (\ref{eq:21}) and (\ref{eq:2}) and using Property 1 yields
\begin{align}
\Delta \dot x_o=-\beta \hat J(q)\hat J^{T}(q)\Delta x_o+Y_k(q,\dot q)\Delta a_k-\hat J(q)s.
\end{align}
By premultiplying $s=\dot q-\dot q_r$ with $\hat J(q)$ and using Property 1, we obtain
\begin{align}
\hat J(q)s=&\dot x+\hat J(q) \hat J^{ T}(q) \left[\gamma(x_o-x_d)\right]+Y_k(q,\dot q)\Delta a_k.\end{align}
Therefore, at the outer loop, we obtain
\begin{align}
\label{eq:28}
\Delta \dot x_o=&-\beta \hat J(q)\hat J^{T}(q)\Delta x_o+Y_k(q,\dot q)\Delta a_k-\hat J(q)s
\\
\label{eq:29}
\dot x=&-\hat J(q) \hat J^{ T}(q) \left[\gamma(x_o-x_d)\right]-Y_k(q,\dot q)\Delta a_k+\hat J(q)s.
\end{align}

We are presently ready to formulate the following theorem.

\emph{Theorem 2: } Suppose that $\hat{\mathcal K}_I$ is uniformly positive definite and $\hat J(q)$ has full row rank, and let the controller parameters $\beta$ and $\gamma$ be chosen such that
\be
\label{eq:30}
\beta>4\gamma/9.
\ee
The adaptive outer loop controller given by (\ref{eq:23}), (\ref{eq:12}), (\ref{eq:21}), (\ref{eq:24}), (\ref{eq:25}), (\ref{eq:16}), and (\ref{eq:17}) with $\dot q_r$ being given as (\ref{eq:22}) for the robotic system (\ref{eq:2}) and (\ref{eq:3}) under the inner PI controller (\ref{eq:6}) ensures the stability of the system and convergence of the task-space position error, i.e., $\Delta x\to 0$ as $t\to\infty$.

\emph{Proof:} Substituting (\ref{eq:6}), (\ref{eq:23}), and (\ref{eq:12}) into (\ref{eq:3}) and using Property 4 gives
\begin{align}
\label{eq:31}
M(q)\dot s+C(q,\dot q)s=&-(K^\ast+B) s-KK_I\left[\int_0^t s(\sigma)d\sigma+\delta_0\right]\nn\\
&+{\rm diag}[Y_d(q,\dot q,\dot q_r,\ddot q_r)\hat a_d]K^\ast \Delta w \nn\\
&-K^\ast{\rm diag}[q_c-q_r]\Delta w_I\nn\\
&+Y_d(q,\dot q,\dot q_r,\ddot q_r)\Delta a_d.
\end{align}
Consider the Lyapunov-like function candidate
\begin{align}
V^\ast=&\frac{1}{2}s^T M(q)s+\frac{1}{2}\left[\int_0^t s(\sigma)d\sigma+\delta_0\right]^T K K_I\nn\\
&\times\left[\int_0^t s(\sigma)d\sigma+\delta_0\right]+\frac{1}{2}\Delta w^T \Lambda^{-1}K^\ast \Delta w\nn\\
&+\frac{1}{2}\Delta a_d^T \Gamma_d^{-1}\Delta a_d+\frac{1}{2}\Delta w_I^T \Lambda_I^{-1}K^\ast \Delta w_I
\end{align}
whose derivative with respect to time along the trajectories of (\ref{eq:31}), (\ref{eq:25}), (\ref{eq:16}), and (\ref{eq:17}) can be written as
\be
\dot V^\ast=-s^T (K^\ast +B)s\le 0
\ee
where we have used Property 3. This directly gives the conclusion that $s\in{\mathcal L}_2\cap {\mathcal L}_\infty$, $\int_0^t s(r)dr\in{\mathcal L}_\infty$, $\hat w\in {\mathcal L}_\infty$, $\hat a_d\in{\mathcal L}_\infty$, and $\hat w_I\in{\mathcal L}_\infty$. Then, there exists a positive constant $\ell_M$ such that $\int_0^t s^T(\sigma)s(\sigma)d\sigma\le \ell_M$, $\forall t\ge 0$. Let us now consider the following quasi-Lyapunov function candidate
\begin{align}
V^{\ast\ast}=&\frac{1}{2}\Delta x_o^T \Delta x_o+\frac{1}{2}\Delta x^T \Delta x+\frac{1}{2}\Delta a_k^T\Gamma_k^{-1}\Delta a_k\nn\\
&+\left(\frac{1}{\beta}+\frac{1}{\gamma}\right)\Big[\ell_M-\int_0^t s^T(\sigma)s(\sigma)d\sigma\Big]
\end{align}
with the choice of the last term following the typical practice (see, e.g., \cite[p.~118]{Lozano2000_Book}), and the derivative of $V^{\ast\ast}$ along the trajectories of (\ref{eq:28}), (\ref{eq:29}), and (\ref{eq:24}) can be written as
\begin{align}
\label{eq:35}
\dot V^{\ast\ast}=&-\beta \Delta x_o^T \hat J(q)\hat J^T(q)\Delta x_o\nn \\
&-\Delta x_o^T \hat J(q)s-\gamma\Delta x^T \hat J(q)\hat J^T(q)\Delta x_o\nn\\
&-\gamma\Delta x^T \hat J(q)\hat J^T(q)\Delta x+\Delta x^T \hat J(q) s-\left(\frac{1}{\beta}+\frac{1}{\gamma}\right)s^T s.\end{align}
Using the following results derived from the standard basic inequalities
\begin{align}
&\Delta x_o^T\hat J(q)s\le \frac{\beta}{4}\Delta x_o^T\hat J(q)\hat J^T(q)\Delta x_o+\frac{1}{\beta}s^T s\\
&\Delta x^T \hat J(q) s\le \frac{\gamma}{4}\Delta x^T \hat J(q) \hat J^T(q)\Delta x+\frac{1}{\gamma}s^T s
\end{align}
we obtain from (\ref{eq:35}) that
\begin{align}
\dot V^{\ast\ast}\le&- \frac{3\beta}{4}\Delta x_o^T \hat J(q)\hat J^T(q)\Delta x_o\nn\\
&-\gamma\Delta x^T \hat J(q)\hat J^T(q)\Delta x_o-\frac{3\gamma}{4}\Delta x^T \hat J(q)\hat J^T(q)\Delta x\nn\\
\le&-\begin{bmatrix}\hat J^T(q)\Delta x_o\\
\hat J^T(q)\Delta x\end{bmatrix}^T\underbrace{\begin{bmatrix}(3\beta/4) I_n & (\gamma/2) I_n\\
(\gamma/2) I_n & (3\gamma/4)I_n\end{bmatrix}}_Q\begin{bmatrix}\hat J^T(q)\Delta x_o\\
\hat J^T(q)\Delta x\end{bmatrix}\le 0
\end{align}
due to the positive definiteness of $Q$ in the case that $\beta>4\gamma/9$, where $I_n$ is the $n\times n$ identity matrix. Then using similar procedures as in the proof of Theorem 1, we can show the stability of the system [regardless of the estimated Jacobian matrix $\hat J(q)$] and the convergence of $\Delta x$ and $\Delta x_o$. \hfill{\small $\blacksquare$}

\emph{Remark 4:} The existing adaptive (or robust) task-space regulation algorithms either assume the exact knowledge of the gravitational torques \cite{Cheah2006_IJRR,Liu2006_TRO}, or require the careful choice of the controller parameters \cite{Cheah2003_TRA,Cheah2010_AUT,Liang2014_ASJC}, or encounter the overparametrization problem \cite{Liang2014_ASJC}. Furthermore, many adaptive visual tracking schemes (e.g., \cite{Cheah2006_IJRR,Wang2010_TCST,Wang2015_AUT,Leite2011_IFAC}) have also been proposed, yet the necessity of investigating task-space regulation algorithms is due to the consideration that the choice of specific controllers should take into account the properties of the specific tasks. It is well accepted that given a specific task, the control law should be as (computationally) simple as possible; while the adaptive tracking controllers can also achieve the regulation of the task-space position to the desired one (constant), it is not cost-effective to rely on such kind of complexity (usually involves the inverse of the estimated Jacobian and the singularity issues) for regulation tasks. The two adaptive regulation controllers presented here rule out the limitations of the above results, and in addition the proposed controllers can be applied to robotic systems with an inner/outer loop structure (e.g., most industrial/commercial robotic systems) that have an unmodifiable joint servoing controller but admit the design of the joint velocity (or position) command, benefiting from the dynamic feedback design and the use of the adaptively scaled dynamic compensation action.

\subsection{Extension to Task-Space Tracking}

In the case of task-space tracking, a feedforward action needs to be introduced in the definition of the joint reference velocity. Specifically
we define $\dot q_r$ as
\be
\label{eq:39}
\dot q_r=\underbrace{\hat J^T(q)[\hat J(q)\hat J^T(q)]^{-1}\dot x_d}_\text{feeforward}\underbrace{-\gamma \hat J^T(q)(x_o-x_d)}_\text{feedback}
\ee
where the use of the generalized inverse of $\hat J(q)$ follows the typical practice. The definition given by (\ref{eq:39}) extends the one in \cite{Wang2016_TAC} to address the case of no task-space velocity measurement. The interesting point here is that both the inverse and transpose of $\hat J(q)$ are incorporated, and one is for introducing a feedforward action and the other for introducing a feedback action. This is in contrast to most existing task-space control algorithms that rely on the use of the inverse of the (estimated) Jacobian matrix to exert both the feedforward and feedback actions (see, e.g., \cite{Slotine1987_IJRR,Niemeyer1991_IJRR,Wang2015_AUT}). As can be clearly observed, once the desired task-space velocity becomes zero, $\dot q_r$ in (\ref{eq:39}) reduces to the one defined by (\ref{eq:22}), and this means that the regulation and tracking cases are unified.

\emph{Theorem 3:} Suppose that $\hat{\mathcal K}_I$ is uniformly positive definite and $\hat J(q)$ has full row rank, and let the controller parameters $\beta$ and $\gamma$ be chosen such that
\be
\beta>4\gamma/9.
\ee
The adaptive outer loop controller given by (\ref{eq:23}), (\ref{eq:12}), (\ref{eq:21}), (\ref{eq:24}), (\ref{eq:25}), (\ref{eq:16}), and (\ref{eq:17}) with $\dot q_r$ being given as (\ref{eq:39}) for the robotic system (\ref{eq:2}) and (\ref{eq:3}) under the inner PI controller (\ref{eq:6}) ensures the stability of the system and convergence of the task-space tracking errors, i.e., $\Delta x\to 0$ and $\Delta \dot x\to 0$ as $t\to\infty$.

The proof of Theorem 3 can be straightforwardly completed based on that of Theorem 2, and the major difference lies in the fact that equation (\ref{eq:29}) for the case of regulation problem now becomes
\begin{align}
\label{eq:41}
\Delta\dot x=-\hat J(q) \hat J^{ T}(q) \left[\gamma(x_o-x_d)\right]-Y_k(q,\dot q)\Delta a_k+\hat J(q)s.
\end{align}
By an analysis of the system given by (\ref{eq:28}), (\ref{eq:41}), and (\ref{eq:24}), we can derive the stability of the system and convergence of the task-space tracking errors.

\emph{Remark 5:} The filter-based adaptive regulation algorithm is computationally simpler in comparison with the observer-based one. But the main issue of the filter-based algorithm is that it is difficult to quantitatively evaluate the performance. Here the observer-based algorithm is extended to cover the case of task-space tracking by additionally introducing feedforward based on the generalized inverse of the estimated Jacobian matrix [see (\ref{eq:39})]. The extension of the filter-based algorithm to realize the task-space tracking can be completed in a similar way.

\subsection{Task-Space Adaptive Control With an Inner PID Position Controller}

We here investigate another case that the low-level controller takes the PID position control action, i.e.,
\be
\label{eq:42}
u=-K_D (\dot q-\dot q_c)-K_P (q-q_c)-K_I \int_0^t [q(\sigma)-q_c(\sigma)]d\sigma
\ee
where $K_D$, $K_P$, and $K_I$ are the derivative, proportional, and integral gains (diagonal and positive definite), respectively. In this case, to ensure the stability and convergence of the robotic system, we need to make some modifications. We take the filter-based adaptive regulation control in Sec. III-A as an illustrating example and the other controllers can be similarly formulated. Specifically, we define two quantities below
\begin{align}
\dot q_r^\ast=&\dot q_r-K_c(q-q_r)\\
\ddot q_r^\ast=&\ddot q_r-K_c(\dot q-\dot q_r)
\end{align}
with $q_r$ being generated by {
\be
\label{eq:45}
\dot q_r=-K_c q_r-K_2 y+K_c q
\ee where $K_c$ is a diagonal positive definite matrix. Let ${\mathcal K}_P=K_D^{-1}K_P$ and $\mathcal {K}_I=K_D^{-1}K_I$ and denote by $\hat {\mathcal K}_P$ and $\hat{\mathcal K}_I$ the estimate of ${\mathcal K}_P$ and that of ${\mathcal K}_I$, respectively, which are specifically written as
\begin{align}
\hat{\mathcal K}_P=&{\rm diag}[\hat w_P]\\
\hat{\mathcal K}_I=&{\rm diag}[\hat w_I]
\end{align}
with $\hat w_P$ and $\hat w_I$ being $n$-dimensional vectors.
 The joint velocity command for the low-level PID controller is defined as
\begin{align}
\label{eq:48}
\dot q_c&+\hat {\mathcal K}_P q_c+\hat {\mathcal K}_I\int_0^t \left[q_c(\sigma)-q_r(\sigma)\right]d\sigma\nn\\
=&\dot q_r^\ast+\hat {\mathcal K}_P q_r+{\rm diag}[\hat w_i,i=1,\dots,n]\nn\\
  &\times\left[-\alpha \hat J^{T}(q)\Delta x+Y_d(q,\dot q,\dot q_r^\ast,\ddot q_r^\ast) \hat a_d\right]. \end{align}

  The adaptation laws for $\hat w$, $\hat a_d$, $\hat w_P$, and $\hat w_I$ are now given as
\begin{align}
\label{eq:49}
&\dot{\hat w}=-\Lambda {\rm diag}[-\alpha \hat J^{T}(q)\Delta x+Y_d(q,\dot q,\dot q_r^\ast,\ddot q_r^\ast)\hat a_d]\xi
\\
\label{eq:50}
&\dot{\hat a}_d=-\Gamma_d Y_d^T(q,\dot q,\dot q_r^\ast,\ddot q_r^\ast) \xi\\
\label{eq:51}
&\dot{\hat {w}}_P=\Lambda_P {\rm diag}[q_c-q_r]\xi\\
\label{eq:52}
&\dot{\hat {w}}_I=\Lambda_I {\rm diag}\left[\int_0^t[q_c(\sigma)-q_r(\sigma)]d\sigma\right]\xi
\end{align}
with
\be\xi=\dot q-\dot q_r^\ast=s+K_c\left[\int_0^t s(\sigma)d\sigma+\delta_0\right],\ee
and $\Lambda_P$ and $\Lambda_I$ being diagonal positive definite matrices, and the adaptation law for $\hat a_k$ is still the same as (\ref{eq:14}).

With these modifications and using the fact that $q-q_r=\int_0^t s(\sigma)d\sigma+\delta_0$, equation (\ref{eq:18}) becomes
\begin{align}
\label{eq:54}
M&(q)\dot \xi+C(q,\dot q)\xi\nn\\
=&-(\bar K^\ast+B) \xi-\alpha \hat J^{ T}(q)\Delta x\nn\\
&-KK_P(q-q_r)-KK_I \int_0^t [q(\sigma)-q_r(\sigma)]d\sigma\nn\\
&+{\rm diag}[-\alpha \hat J^{T}(q)\Delta x+Y_d(q,\dot q,\dot q_r^\ast,\ddot q_r^\ast)\hat a_d]\bar K^\ast \Delta w \nn\\
&-\bar K^\ast{\rm diag}[q_c-q_r]\Delta w_P-\bar K^\ast{\rm diag}\left[\int_0^t[q_c(\sigma)-q_r(\sigma)]d\sigma\right]\Delta w_I\nn\\
&+Y_d(q,\dot q,\dot q_r^\ast,\ddot q_r^\ast)\Delta a_d\nn\\
=&-(\bar K^\ast+B) \xi-\alpha \hat J^{ T}(q)\Delta x-(KK_P-K K_I K_c^{-1})(q-q_r)\nn\\
&-KK_I K_c^{-1}\left[q-q_r+K_c\int_0^t [q(\sigma)-q_r(\sigma)]d\sigma\right]\nn\\
&+{\rm diag}[-\alpha \hat J^{T}(q)\Delta x+Y_d(q,\dot q,\dot q_r^\ast,\ddot q_r^\ast)\hat a_d]\bar K^\ast \Delta w \nn\\
&-\bar K^\ast{\rm diag}[q_c-q_r]\Delta w_P-\bar K^\ast{\rm diag}\left[\int_0^t[q_c(\sigma)-q_r(\sigma)]d\sigma\right]\Delta w_I\nn\\
&+Y_d(q,\dot q,\dot q_r^\ast,\ddot q_r^\ast)\Delta a_d\nn\\
=&-(\bar K^\ast+B) \xi-\alpha \hat J^{ T}(q)\Delta x\nn\\
&-(KK_P-K K_I K_c^{-1})\left[\int_0^t s(\sigma)d\sigma+\delta_0\right]\nn\\
&-KK_I K_c^{-1}\left[\int_0^t \xi(\sigma)d\sigma+\delta_0\right]\nn\\
&+{\rm diag}[-\alpha \hat J^{T}(q)\Delta x+Y_d(q,\dot q,\dot q_r^\ast,\ddot q_r^\ast)\hat a_d]\bar K^\ast \Delta w \nn\\
&-\bar K^\ast{\rm diag}[q_c-q_r]\Delta w_P-\bar K^\ast{\rm diag}\left[\int_0^t[q_c(\sigma)-q_r(\sigma)]d\sigma\right]\Delta w_I\nn\\
&+Y_d(q,\dot q,\dot q_r^\ast,\ddot q_r^\ast)\Delta a_d
\end{align}
where $\bar K^\ast=K K_D$. The above equation can further be written as
\begin{align}
\label{eq:55}
M&(q)\dot \xi+C(q,\dot q)\xi\nn\\
=&{\rm diag}[-\alpha \hat J^{T}(q)\Delta x+Y_d(q,\dot q,\dot q_r^\ast,\ddot q_r^\ast)\hat a_d]\bar K^\ast \Delta w \nn\\
&-\bar K^\ast{\rm diag}[q_c-q_r]\Delta w_P-\bar K^\ast{\rm diag}\left[\int_0^t[q_c(\sigma)-q_r(\sigma)]d\sigma\right]\Delta w_I\nn\\
&+Y_d(q,\dot q,\dot q_r^\ast,\ddot q_r^\ast)\Delta a_d-\alpha \hat J^{T}(q)\Delta x\nn\\
&-K\Big[(K_D+K^{-1}B)s\nn\\
&+[(K_D+K^{-1}B)K_c+(K_P- K_I K_c^{-1})]\left(\int_0^t s(\sigma)d\sigma+\delta_0\right)\Big]\nn\\
&-KK_I K_c^{-1}\left[\int_0^t \xi(\sigma)d\sigma+\delta_0\right].
\end{align}
Consider the Lyapunov function candidate
\begin{align}
\label{eq:56}
V=&\alpha\Big(\frac{1}{2}\Delta x^T \Delta x+\frac{1}{2}y^TK_2 K_1^{-1} y+\frac{1}{2}\Delta a_k^T \Gamma_k^{-1}\Delta a_k\Big)\nn\\
&+\frac{1}{2}\xi^T M(q)\xi+\frac{1}{2}\left[\int_0^t \xi(\sigma)d\sigma+\delta_0\right]^T K K_I \left[\int_0^t \xi(\sigma)d\sigma+\delta_0\right]\nn\\
&+\frac{1}{2}\left[\int_0^t s(\sigma)d\sigma+\delta_0\right]^T(K{\mathcal M}+KK_cK_D+K_cB)\nn\\
&\times\left[\int_0^t s(\sigma)d\sigma+\delta_0\right]+\frac{1}{2}\Delta w^T \Lambda^{-1}\bar K^\ast \Delta w+\frac{1}{2}\Delta a_d^T \Gamma_d^{-1}\Delta a_d\nn\\
&+\frac{1}{2}\Delta w_P^T \Lambda_P^{-1}\bar K^\ast \Delta w_P+\frac{1}{2}\Delta w_I^T \Lambda_I^{-1}\bar K^\ast \Delta w_I
\end{align}
where \be
\label{eq:57}
{\mathcal M}=(K_D+K^{-1}B)K_c+K_P-K_IK_c^{-1}.
\ee By suitably choosing $K_c$, we can ensure that ${\mathcal M}$ is positive semidefinite. The derivative of $V$ can be written as
\begin{align}
\dot V=&-\alpha y^T K_2 y-s^T (\bar K^\ast+B)s\nn\\
&-\left[\int_0^t s(\sigma)d\sigma+\delta_0\right]^T K{\mathcal M}K_c\left[\int_0^t s(\sigma)d\sigma+\delta_0\right]\le 0.
\end{align}

\emph{Theorem 4:} Suppose that $\hat{\mathcal K}_P$ and $\hat{\mathcal K}_I$ evolve such that the following system
\be
\label{eq:a1}
\ddot z+\hat{\mathcal K}_P\dot z+\hat{\mathcal K}_I z=0
\ee
with $z\in R^n$ is uniformly exponentially stable and that $\hat J(q)$ has full row rank, and choose the matrix $K_c$ such that ${\mathcal M}$ given by (\ref{eq:57}) is positive semidefinite. Then the adaptive outer loop controller given by (\ref{eq:48}), (\ref{eq:12}), (\ref{eq:7}), (\ref{eq:45}), (\ref{eq:49}), (\ref{eq:50}), (\ref{eq:51}), (\ref{eq:52}), and (\ref{eq:14}) for the robotic system (\ref{eq:2}) and (\ref{eq:3}) under the inner PID position controller (\ref{eq:42}) ensures the stability of the system and convergence of the task-space position error, i.e., $\Delta x\to 0$ as $t\to\infty$.

The proof of Theorem 4 can be completed by following similar steps as in the proof of Theorem 1.

\emph{Remark 6:} \begin{enumerate}

\item One key issue in the case of using a low-level PID controller is the choice of $K_c$, and obviously large enough $K_c$ can ensure that $\mathcal M$ is positive semidefinite. The remaining thing is how to determine the gain $K_c$. In practice, since the damping matrix $B$ may be quite small, we thus neglect it and this implies that $k_{c,ii}$ (i.e., the $i$-th diagonal entry of $K_c$) should satisfy
\be
\label{eq:58}
k_{c,ii}\ge\frac{2k_{I,ii}}{k_{P,ii}+\sqrt{k_{P,ii}^2+4 k_{I,ii}k_{D,ii}}}, \forall i=1,\dots,n
\ee
where $k_{D,ii}$ is the $i$-th diagonal entry of $K_D$, $k_{P,ii}$ is the $i$-th diagonal entry of $K_P$, and $k_{I,ii}$ is the $i$-th diagonal entry of $K_I$. Here we face the similar situation as the control engineers of the robot production company. The control engineers are usually careful about the choice of the integral gain and large gain may cause instability; a trade-off has to be made between the attenuation of constant disturbances and the stability margin of the control system. In practice, $k_{P,ii}$ and $k_{D,ii}$ are possibly/generally chosen to be not less than $k_{I,ii}$, and in this case, we can simply choose $k_{c,ii}$ as $k_{c,ii}\ge(\sqrt{5}-1)/2$, $\forall i=1,\dots,n$.

\item An important issue in proving Theorem 4 is to clarify the input-output properties of
 the following system [derived from (\ref{eq:48})]
\begin{align}
\label{eq:a2}
\dot q_c&-\dot q_r+\hat{\mathcal K}_P(q_c-q_r)+\hat{\mathcal K}_I\int_0^t [q_c(\sigma)-q_r(\sigma)]d\sigma\nn\\
=&{\rm diag}[\hat w_i,i=1,\dots,n]\left[-\alpha \hat J^{T}(q)\Delta x+Y_d(q,\dot q,\dot q_r^\ast,\ddot q_r^\ast) \hat a_d\right]\nn\\
&-K_c(q-q_r)
  \end{align}
where the right side is bounded. The main issue can now be reduced to investigating the stability of the linear time-varying system (\ref{eq:a1}). By the assumption that the system (\ref{eq:a1}) is uniformly exponentially stable, we can directly obtain from (\ref{eq:a2})
that $\int_0^t[q_c(\sigma)-q_r(\sigma)]d\sigma$, $q_c-q_r$, and $\dot q_c-\dot q_r$ are bounded. Then the boundedness of $\ddot q$
can be ensured.

On the other hand, by the standard projection algorithms \cite{Ioannou1996_Book}, we can conveniently ensure that $\hat{\mathcal K}_P$ and
$\hat{\mathcal K}_I$ are uniformly positive. In addition, the boundedness of $\hat{\mathcal K}_P$ and
$\hat{\mathcal K}_I$ is a direct consequence of that of the Lyapunov function
candidate given by (\ref{eq:56}). But even under these two conditions, we still cannot ensure the uniform exponential stability of (\ref{eq:a1}) [which is a sufficient condition to ensure the uniform bounded-input bounded-output stability of (\ref{eq:a2}) according
to the standard linear system theory]
since the coefficient matrices are time-varying. In practice, we may slow down the adaptation to ${\mathcal K}_P$ and ${\mathcal K}_I$ (i.e., slow down the variation of $\hat{\mathcal K}_P$
and $\hat{\mathcal K}_I$) so that the system (\ref{eq:a1}) is a slowly time-varying (or quasi-time-invariant) linear system.

\end{enumerate}

\section{Generalizations and Further Discussions}

Let us now discuss the proposed framework in other closely related topics concerning adaptive robot control.

\subsection{Direct Adaptation}

The first result that we would like to discuss is the well-known Slotine and Li adaptive controller \cite{Slotine1987_IJRR} and the result there is presented in the context of open joint torque control. If we redefine $\dot q_r$ in (\ref{eq:22}) as
\be
\label{eq:60}
\dot q_r=\dot q_d-\bar\alpha (q-q_d)
\ee with $q_d \in R^n$ being the desired joint position and $\bar\alpha$ a positive design constant, then the adaptive controller given by
(\ref{eq:23}), (\ref{eq:12}), (\ref{eq:15}), (\ref{eq:16}), and (\ref{eq:17}) with $\dot q_r$ being defined by (\ref{eq:60}) becomes an outer loop (``applicable") version of Slotine and Li adaptive controller suitable for practical robotic systems (without opening the torque control module) with an unmodifiable inner PI velocity control loop (in the case of an inner PID position controller, its outer loop version can be similarly developed by following the steps in Sec. III-D). In this case of the joint-space position tracking, one can easily show that the position command $q_c=q_d$ is also qualified for ensuring the stability of the robotic system and convergence of the joint tracking errors.

\subsection{Composite Adaptation}

The implementation of the standard composition adaptation algorithm given in \cite{Slotine1989_Aut} in the framework of inner/outer controller structure seems not straightforward and we need to ensure that no additional number of parameters appear in the filtered dynamic model (i.e., avoiding the overparameterization). For this purpose, we rewrite the dynamics (\ref{eq:3}) as
\begin{align}
&{\rm diag}[w_i,i=1,\dots,n] [M(q)\ddot q+C(q,\dot q)\dot q+B\dot q+g(q)]\nn\\
&=K_P^{-1}u
=-(\dot q-\dot q_c)-K_P^{-1}K_I\left(q-q_c\right)
\end{align}
and by using the filtering technique in \cite{Slotine1989_Aut}, we then have the following equation without involving joint acceleration measurement
\begin{align}
&{\rm diag}[w_i,i=1,\dots,n] Y_f(q,\dot q, t) a_d\nn\\
&=u_f^\ast-{\rm diag}[w_{I,i},i=1,\dots,n]h_f
\end{align}
where $w_i=k_i^{\ast -1}$, $w_{I,i}={k_{P,ii}^{-1}}k_{I,ii}$, $i=1,\dots,n$, $Y_f(q,\dot q,t)=\frac{\lambda_f}{p+\lambda_f} Y_d(q,\dot q,\dot q,\ddot q)$, $u_f^\ast=-\frac{\lambda_f}{p+\lambda_f}(\dot q-\dot q_c)$, and $h_f=\frac{\lambda_f}{p+\lambda_f}\left(q-q_c\right)$ with $p$ and $\lambda_f>0$ being the Laplace variable and the filter parameter, respectively. Let
\be
\hat u_f^\ast={\rm diag}[\hat w] Y_f \hat a_d+{\rm diag}[h_f]\hat {w}_I
\ee
where $\hat w_I$ is the estimate of $w_I=[w_{I,1},\dots,w_{I,n}]^T$. Then we define a prediction error
\begin{align}
\label{eq:64}
e_f=&\hat u_f^\ast-u_f^\ast\nn\\
=&{\rm diag}[w] Y_f\Delta a_d+{\rm diag}[Y_f \hat a_d]\Delta w+{\rm diag}[h_f]\Delta {w}_I
\end{align}
and the composite adaptive version of (\ref{eq:15}), (\ref{eq:16}), and (\ref{eq:17}) is given as
\begin{align}
\label{eq:65}
\dot{\hat w}=&-\Lambda ({\rm diag}[Y_d(q,\dot q,\dot q_r,\ddot q_r) \hat a_d]s+\gamma_0{\rm diag}[Y_f \hat a_d]e_f)\\
\label{eq:66}
\dot{\hat a}_d=&-\Gamma_d (Y_d^T(q,\dot q,\dot q_r,\ddot q_r)s +\gamma_0 Y_f^T e_f)\\
\label{eq:67}
\dot{{\hat w}}_I=&\Lambda_I({\rm diag}[q_c-q_r]s-\gamma_0 {\rm diag}[h_f] e_f)
\end{align}
where $\gamma_0$ is a positive design constant.

\emph{Remark 7:} The interesting and also distinguished point here is that the prediction error $e_f$ given by (\ref{eq:64}) contains the unknown coefficient matrix ${\rm diag}[w]$, due to which the regressor matrix is actually only partially known. This motivates us to wonder whether or not the stability of the closed-loop robotic system can still be guaranteed under the composite adaptation here. 

Consider the nonnegative function
\be
\label{eq:68}
V_1=(1/2)[\Delta w^T \Lambda^{-1} K^\ast \Delta w+\Delta a_d^T \Gamma_d^{-1}\Delta a_d+\Delta w_I^T \Lambda_I^{-1}K^\ast\Delta w_I]
\ee
whose derivative along (\ref{eq:65}), (\ref{eq:66}), and (\ref{eq:67}) can be written as
\begin{align}
\dot V_1=&-\Delta w^T K^\ast {\rm diag}[Y_d(q,\dot q,\dot q_r,\ddot q_r) \hat a_d]s-\Delta a_d^T Y_d^T (q,\dot q,\dot q_r,\ddot q_r)s\nn\\
&+\Delta w_I^TK^\ast{\rm diag}[q_c-q_r]s-\gamma_0 (\Delta w ^T K^\ast {\rm diag}[Y_f \hat a_d]e_f\nn\\
&+\Delta a_d^T Y_f^T e_f+\Delta w_h^T K^\ast{\rm diag}[h_f]e_f)\nn\\
=&{-\Delta w^T K^\ast {\rm diag}[Y_d(q,\dot q,\dot q_r,\ddot q_r) \hat a_d]s-\Delta a_d^T Y_d^T(q,\dot q,\dot q_r,\ddot q_r) s}\nn\\
&+\Delta w_I^TK^\ast{\rm diag}[q_c-q_r]s-\gamma_0 e_f^T K^\ast\nn\\
&\times (\underbrace{{\rm diag}[Y_f \hat a_d]\Delta w+{\rm diag}[w]Y_f\Delta a_d+{\rm diag}[h_f]\Delta w_I}_{e_f})
\end{align}
where the first three terms are used to compensate for the indefinite terms due to the parametric uncertainty, and in this way, the stability is ensured. The key point here is to exploit the independent nature of the joint processors which means that $K^\ast$ is diagonal. The other versions of composite adaptation, e.g., BGF composite adaptation and CF composite adaptation (see, e.g., \cite{Slotine1989_Aut}) can also be used so that smoother parameter adaptation and better tracking performance can be achieved. Specifically, the CF composite adaptation laws can be given as
\be
\begin{cases}
\dot{\hat w}=-\Lambda ({\rm diag}[Y_d(q,\dot q,\dot q_r,\ddot q_r) \hat a_d]s+\gamma_0{\rm diag}[Y_f \hat a_d]e_f)\\
\dot \Lambda=\lambda_1(\Lambda-\Lambda \bar\Lambda^{-1} \Lambda)-\gamma_0\Lambda ({\rm diag}[Y_f \hat a_d])^2 \Lambda\\
\dot{\hat a}_d=-\Gamma_d (Y_d^T(q,\dot q,\dot q_r,\ddot q_r)s +\gamma_0 Y_f^T e_f)\\
\dot \Gamma_d=\lambda_2(\Gamma_d-\Gamma_d \bar\Gamma_d^{-1} \Gamma_d)-\gamma_0\Gamma_d Y_f^T Y_f \Gamma_d\\
\dot{{\hat w}}_I=\Lambda_I({\rm diag}[q_c-q_r]s-\gamma_0 {\rm diag}[h_f] e_f)\\
\dot \Lambda_I=\lambda_3(\Lambda_I-\Lambda_I \bar\Lambda_I^{-1} \Lambda_I)-\gamma_0\Lambda_I ({\rm diag}[h_f])^2 \Lambda_I
\end{cases}
\ee
where $\lambda_1$, $\lambda_2$, and $\lambda_3$ are strictly positive forgetting factors (time-varying or constant), and $\bar\Lambda$, $\bar \Gamma_d$, and $\bar\Lambda_I$ denote the upper bounds of $\Lambda$, $\Gamma_d$, and $\Lambda_I$, respectively. Note that $\Lambda(0)$ and $\Lambda_I(0)$ are chosen as diagonal positive definite matrices satisfying $0<\Lambda(0)\le \bar\Lambda$ and $0<\Lambda_I(0)\le \bar\Lambda_I$, and $\Gamma_d(0)$ can be chosen as a symmetric positive definite matrix satisfying $0<\Gamma_d(0)\le \bar\Gamma_d$, and in this way, it can be shown that $\Lambda(t)$ and $\Lambda_I(t)$ are always diagonal, $\forall t\ge 0$. The derivative of the nonnegative function $V_1$ defined by (\ref{eq:68}) in this case becomes
\begin{align}
\dot V_1=&{-\Delta w^T K^\ast {\rm diag}[Y_d(q,\dot q,\dot q_r,\ddot q_r) \hat a_d]s-\Delta a_d^T Y_d^T(q,\dot q,\dot q_r,\ddot q_r) s}\nn\\
&+\Delta w_I^TK^\ast{\rm diag}[q_c-q_r]s\nn\\
&-\frac{\lambda_1}{2}\Delta w^T(\Lambda^{-1}-\bar\Lambda^{-1})K^\ast\Delta w-\frac{\lambda_2}{2}\Delta a_d^T(\Gamma_d^{-1}-\bar\Gamma_d^{-1})\Delta a_d\nn\\
&-\frac{\lambda_3}{2}\Delta w_I^T(\Lambda_I^{-1}-\bar\Lambda_I^{-1})K^\ast \Delta w_I-\frac{\gamma_0}{2} e_f^T K^\ast e_f.
\end{align}
Obviously, the stability of the system and convergence of the joint tracking errors can be guaranteed.

\subsection{Adaptive Control of Flexible-Joint Manipulators}

The typical result may be the singular-perturbation-based adaptive control approach \cite{Spong1989_SCL,Spong1995_AUT}. Consider a flexible-joint manipulator governed by \cite{Spong1987_ASME}
\be
\label{eq:72}
\begin{cases}
M_0(q)\ddot q+C(q,\dot q)\dot q+B\dot q+g(q)=K_s(\theta-q)\\
D_r \ddot \theta+B_r\dot \theta=K u-K_s(\theta-q)
\end{cases}
\ee
where $\theta\in R^n$ is the rotor position, $K_s$ is the constant, diagonal, and positive definite stiffness matrix, $D_r\in R^{n\times n}$ is the rotor inertia matrix seen from the link side, and $B_r\in R^{n\times n}$ is the damping matrix. In this case, the motor velocity command is defined as
\begin{align}
 \label{eq:73}
 \dot q_c+\hat {\mathcal K}_I q_c=&\dot q_r+\hat {\mathcal K}_I q_r+{\rm diag}[\hat w_i,i=1,\dots,n]\nn\\
  &\times Y_d(q,\dot q,\dot q_r,\ddot q_r) \hat a_d
 \end{align}
 with $\dot q_r$ being defined as
 \begin{align}
 \dot q_r= &\dot q_d-\bar\alpha(q-q_d).
 \end{align}
The adaptation laws for $\hat w$, $\hat a_d$, and $\hat w_I$ are given as
\begin{align}
\label{eq:75}
&\dot{\hat w}=-\Lambda {\rm diag}[Y_d(q,\dot q,\dot q_r,\ddot q_r)\hat a_d]s
\\
\label{eq:76}
&\dot{\hat a}_d=-\Gamma_d Y_d^T(q,\dot q,\dot q_r,\ddot q_r) s\\
\label{eq:77}
&\dot{\hat w}_I=\Lambda_I{\rm diag}[q_c-q_r]s
\end{align}
The low-level PI control action in this case of flexible-joint robots would typically take the form
\be
\label{eq:78}
u=-K_P (\dot \theta-\dot q_c)-K_I \left(\theta-q_c\right).
\ee
Substituting (\ref{eq:78}) into the second equation of (\ref{eq:72}) with some further manipulations gives
\begin{align}
D_r&(\ddot \theta-\ddot q)+B_r(\dot \theta-\dot q)\nn\\
=&-K^\ast (\dot \theta-\dot q_c)-KK_I\left(\theta-q_c\right)\nn\\
&-D_r\ddot q-B_r\dot q-K_s(\theta-q)\end{align}
and the above equation can further be written as
\begin{align}
D_r&(\ddot\theta-\ddot q)+(B_r+K^\ast)(\dot\theta-\dot q)+[(K_s+KK_I)K_s^{-1}]K_s(\theta-q)\nn\\
=&-K^\ast(\dot q-\dot q_c)-KK_I\left(q-q_c\right)-D_r\ddot q-B_r\dot q.
\end{align}
As the fast dynamics becomes settled, i.e., $K_s(\theta-q)$ is quasi-constant, we obtain that
\begin{align}
K_s(\theta-q)=&-K_s(K_s+KK_I)^{-1}\Big\{K^\ast(\dot q-\dot q_c)\nn\\
&+KK_I\left(q-q_c\right)+D_r\ddot q+B_r\dot q\Big\}
\end{align}
and thus the slow dynamics becomes
\begin{align}
[M_0&(q)+K_s^\ast D_r]\ddot q+C(q,\dot q)\dot q+(B+K_s^\ast B_r)\dot q+g(q)\nn\\
=&-K_s^\ast K^\ast\left[(\dot q-\dot q_c)+{\mathcal K}_I\left(q-q_c\right)\right]
\end{align}
where $K_s^\ast=K_s(K_s+KK_I)^{-1}$. The fast/slow-dynamics-based analysis given above is based on \cite{Spong1989_SCL,Spong1995_AUT}, and one can rigorously obtain the stability and convergence of the system by following similar arguments as in \cite{Spong1989_SCL,Spong1995_AUT}.

\emph{Remark 8:} Different from the rigid robot case, the actual value of the scale parameter now satisfies ${\rm diag}[w]=K_s^\ast K^\ast$. Part of the low-level integral action in (\ref{eq:78}) $-K_I(\theta-q)$ is the same as the relative position feedback in \cite{Spong1995_AUT} and its effect is to increase the joint stiffness, providing the possibility of applying the control to manipulators with a relatively low joint stiffness \cite{Spong1995_AUT}.

\emph{Remark 9:} It is interesting to note that the joint velocity and position commands in the flexible joint case remains the same as the rigid joint case. This provides a good understanding and more importantly an effective justification of why most results derived in the case of rigid robots are generally applicable to (not justified in the previous literature though) practical robotic systems with inner/output loop structure (e.g., most industrial/commercial robotic systems), even without the need of any modification (in practice, any robot has certain joint flexibility). In fact, the relative damping suggested in \cite{Spong1989_SCL} is naturally included as applying the scaled-dynamic-compensation versions of most control schemes valid for rigid robots to (flexible-joint) robots in practice. More remarks in terms of the roles of the rotor inertias and joint stiffness are presented in the later simulation.

One may also be interested in deriving a composite adaptive version of the adaptive scheme for flexible-joint manipulators and it shall be feasible by using similar techniques as those for rigid manipulators.

\subsection{Further Discussions}

Here, we take several standard adaptive robot control schemes for illustrating how the scaled dynamic compensation makes them to be qualified adaptive outer loop schemes and further the possible applications to robotic systems with an inner/outer loop structure. It seems hopeful that most adaptive robot control schemes in the literature can be reshaped to be adaptive outer loop schemes by accommodating such modifications.

The inner/outer controller structure basically performs the inner joint servoing much faster and the outer loop relatively slower, and thus the scaled dynamic compensation is actually exerted at a quite lower updating cycle. In the case of fast operating process, this would result in degrading of the performance and even instability at certain extreme cases. The main objective of the study and results presented here is to provide the possibility of exerting dynamic compensation (feedforward) even in the standard setting of industrial robotic systems, of course, under the limit of the operation speed. Once upon a while, direct-drive robots are believed to be promising in taking over the role of the standard robots using gear reduction in that direct-drive robots are much efficient and less influenced by friction and backlash, etc. But this hope advances not so favorably, especially in applications, and the reasons may perhaps be the following:
\begin{itemize}
\item The torque output of direct-drive robots is small and large torque output would require large and heavy joint motors, which, however, are significantly constrained by the weight limit of the manipulator;
\item direct-drive robots, as is typically expected, are torque-based, but torque-based design is relatively risky and not so reliable since all factors are taken into account at the same time and in addition the communication constraint presents a limit of the coupling torque exerting cycle.
\end{itemize}
These unfavorable factors concerning torque-based design and direct-drive robots give rise to the welcome of the inner/outer loop structure in most practical robotic applications and this may still be going in the future. A very recent example is Robonaut 2 \cite{Ahlstrom2013_robonaut} which uses harmonic gear transmission instead of the direct-drive configuration, and the feedforward is shown to be necessary to improve the control accuracy as well as the system response within the range of the torque limit. In this specific example, the joint torque control loop is open just for admitting the injection of feedforward or dynamic control action, but from a long run and for promoting the large-scale production and decreasing the cost, the inner/outer loop structure with the joint control loop sealed may perhaps be more desirable.

\section{Simulation Results}

\subsection{Task-Space Adaptive Control}

Consider a three-DOF manipulator with a tool, as is shown in Fig. 2. Its physical parameters are given in Table I with the labels $1$, $2$, $3$, and $E$ denoting link $1$, $2$, $3$, and the tool, respectively, and the diagonal rotor inertia matrix (seen from the link side) $D_r={\rm diag}[0.6,0.3,0.1]$. The diagonal matrix $B$ and $K$ are set as $B={\rm diag}[0.20,0.15,0.10]$ and $K={\rm diag}[60.0,30.0,10.0]$, respectively. The angle about the axis ${\bf Z}_3$ between the tool and the third link is $\delta=30 \text{ deg}$. The inner joint servoing loop with a PI velocity controller (the case of PID position controller is considered in Sec. V-D) is operated at a high-rate cycle with the sampling period being $0.5 \text{ ms}$, and the outer loop is operated at a low-rate cycle with the sampling period being $20 \text{ ms}$. The gains of the low-level PI controller are set as $K_P=30.0I_3$\footnote{$I_{\ell}$ denotes the $\ell\times \ell$  identity matrix, $\ell=2,3,\dots$.} and $K_I=15.0 I_3$, and this means that the effective PI gains are $K_P^\ast={\rm diag}[1800.0,900.0,300.0]$ and $K_I^\ast={\rm diag}[900.0,450.0,150.0]$.

\begin{figure}
\centering
\begin{minipage}[t]{1.0\linewidth}
\centering
\includegraphics[width=2.5in]{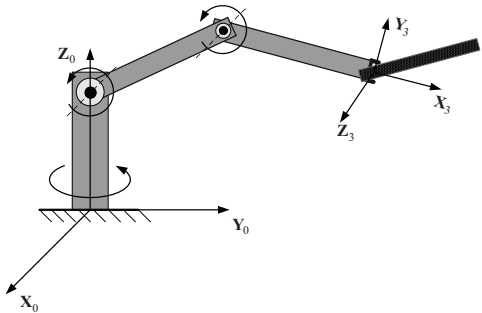}
\caption{Three-DOF manipulator.}
\end{minipage}%
\end{figure}

\begin{table}[!t]
\renewcommand{\arraystretch}{1.5}
\center\caption{Physical parameters of the manipulator}
\begin{tabular}{c c c c c}
\hline
 $i$-th body &$m_i({\rm kg})$&$I_{C,i}^{xx},I_{C,i}^{yy},I_{C,i}^{zz}({\rm kg\cdot m^2})$ &$l_i({\rm m})$&$l_{C,i}({\rm m})$ \\
\hline
1& 1.6000 &   0.4320,   0.0720,  0.4320 &   1.8000&    0.9000 \\
\hline
2& 0.6000&    0.0054,    0.1620,    0.1620 &  1.8000&  0.9000 \\
\hline
3& 0.6000&   0.0054,   0.1620,    0.1620 &   1.8000 &   0.9000 \\
\hline
E&
0.8000&    0.0032,    0.0960,    0.0960&    1.2000&    0.6000 \\
\hline
\end{tabular}
\label{Tab1}
\end{table}

\subsubsection{Regulation problem}

We first perform the simulations of the system under the filter-based and observer-based task-space regulation schemes. The manipulator starts at the configuration $q(0)=[\pi/6,\pi/3,-5\pi/6]^T$ and the corresponding task-space position is $x(0)=[-0.7500,
    1.2990,
    0.5196]^T$. The desired task-space position is set as $x_d=[-1.0,
           2.0,
           0.8]^T$. In the case of using the filter-based scheme, the controller parameters are set as $K_1=60.0 I_3$, $K_2=2.0 I_3$, $\alpha=2.0$, $\Gamma_k=20.0 I_3$, $\Lambda=0.001 I_3$, $\Gamma_d=0.006 I_{15}$, and $\Lambda_I=100.0 I_3$. The initial values of the parameter estimates are chosen as $\hat a_k(0)=[3.0,5.0,2.0]^T$, $\hat w(0)=0_3$, $\hat a_d(0)=0_{15}$, and $\hat w_I(0)=[1,1,1]^T$. Simulation results are shown in Fig. 3, Fig. 4, and Fig. 5, which, respectively, give the task-space position errors, the scale parameter estimates, and the estimate of $w_I$. In the case of using the observer-based scheme, the controller parameters $\beta$ and $\gamma$ are determined as $\beta=1.0$ and $\gamma=1.0$, which obviously satisfy the condition (\ref{eq:30}), and the other controller parameters and the initial parameter estimates are chosen to be the same as those of the filter-based scheme. The task-space position error, the scale parameter estimates, and the estimate of $w_I$ are shown in Fig. 6, Fig. 7, and Fig. 8, respectively.

\begin{figure}
\centering
\begin{minipage}[t]{1.0\linewidth}
\centering
\includegraphics[width=2.6in]{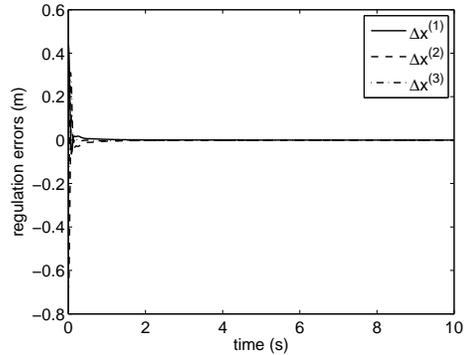}
\caption{Task-space position errors (filter-based control).}\label{fig:side:a}
\end{minipage}%
\end{figure}

\begin{figure}
\centering
\begin{minipage}[t]{1.0\linewidth}
\centering
\includegraphics[width=2.6in]{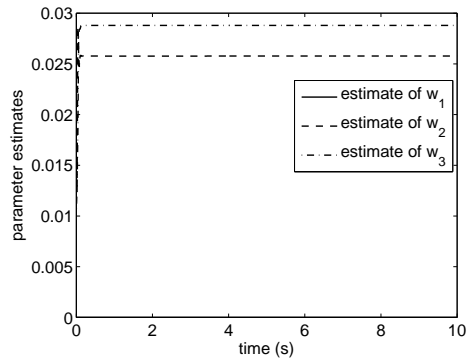}
\caption{Scale parameter estimates (filter-based control).}\label{fig:side:a}
\end{minipage}%
\end{figure}

\begin{figure}
\centering
\begin{minipage}[t]{1.0\linewidth}
\centering
\includegraphics[width=2.6in]{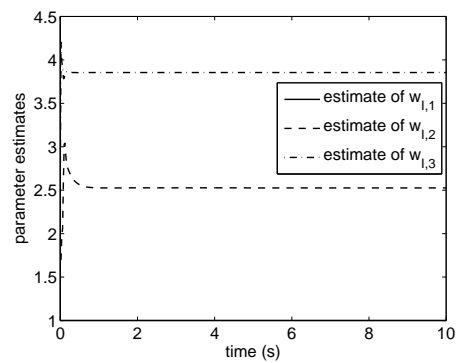}
\caption{Estimate of $w_I$ (filter-based control).}\label{fig:side:a}
\end{minipage}%
\end{figure}

\begin{figure}
\centering
\begin{minipage}[t]{1.0\linewidth}
\centering
\includegraphics[width=2.6in]{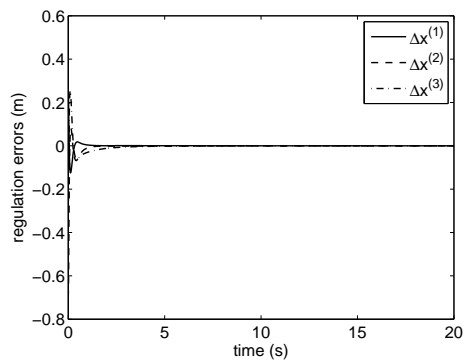}
\caption{Task-space position errors (observer-based control).}\label{fig:side:a}
\end{minipage}%
\end{figure}

\begin{figure}
\centering
\begin{minipage}[t]{1.0\linewidth}
\centering
\includegraphics[width=2.6in]{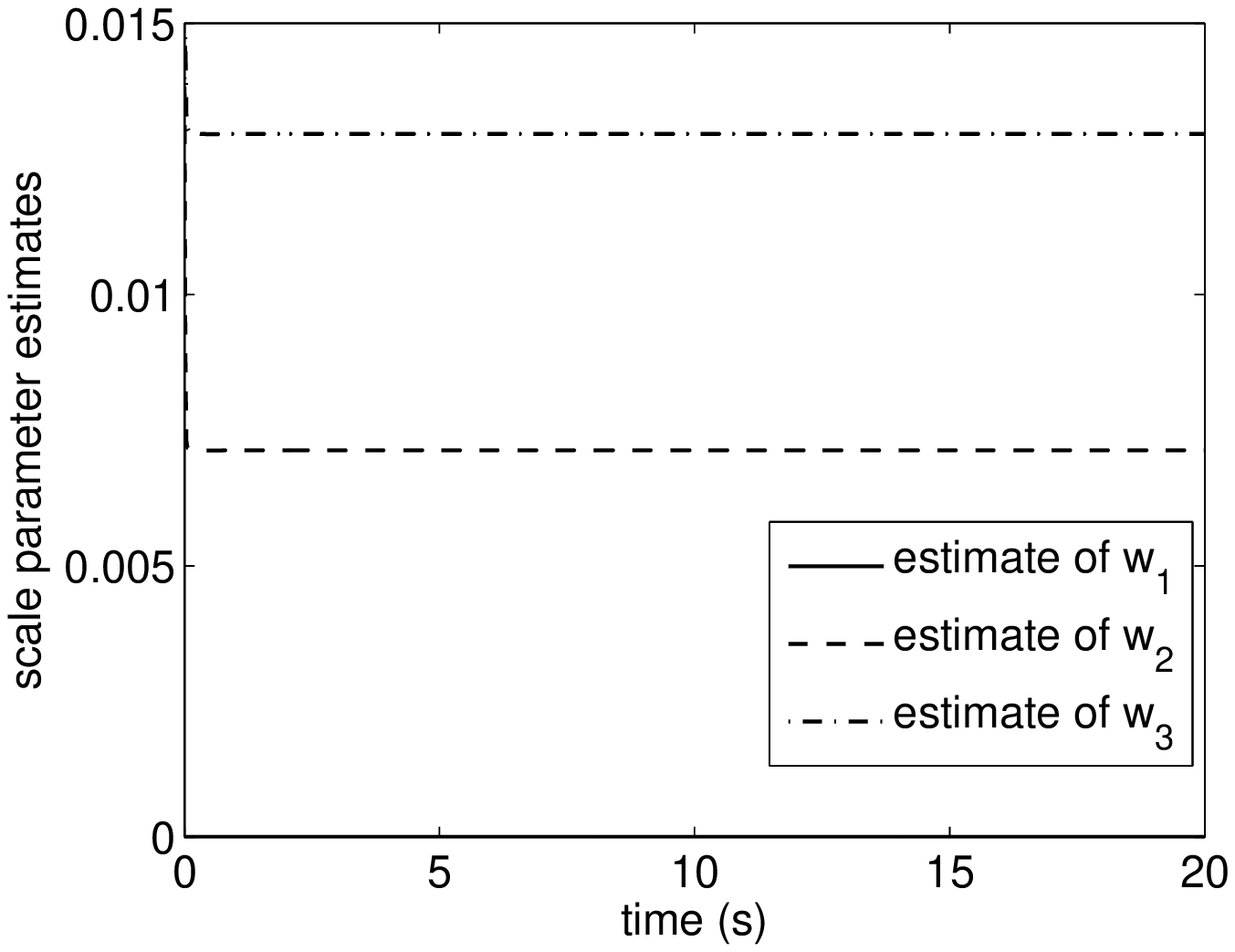}
\caption{Scale parameter estimates (observer-based control).}\label{fig:side:a}
\end{minipage}%
\end{figure}

\begin{figure}
\centering
\begin{minipage}[t]{1.0\linewidth}
\centering
\includegraphics[width=2.6in]{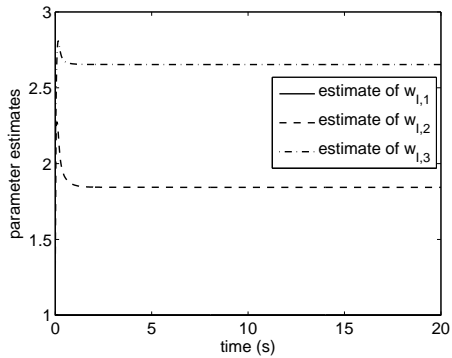}
\caption{Estimate of $w_I$ (observer-based control).}\label{fig:side:a}
\end{minipage}%
\end{figure}

\subsubsection{Tracking problem}

Let us now consider the case of using the observer-based tracking controller given by (\ref{eq:23}), (\ref{eq:12}), (\ref{eq:21}), (\ref{eq:24}), (\ref{eq:25}), (\ref{eq:16}), and (\ref{eq:17}). The desired task-space trajectory is given as $x_d=[-1.0500+0.3\cos(\pi t/3),1.2990+0.3\sin(\pi t/3),0.5196+0.3\sin(\pi t/3)]^T$. The initial configuration of the manipulator is set to be the same as the above. The controller parameters are chosen as $\beta=0.8$, $\gamma=0.8$, $\Gamma_k=200.0 I_3$, $\Lambda=6.0 I_3$, $\Gamma_d=27.0 I_{15}$, and $\Lambda_I=100.0I_3$. The task-space position tracking errors, the scale parameter estimates, and the estimate of $w_I$ are shown in Fig. 9, Fig. 10, and Fig. 11, respectively. The gains are increased, in comparison with the the case of regulation, and this is feasible since in the tracking problem here, the desired trajectory starts at the current position of the manipulator.

\begin{figure}
\centering
\begin{minipage}[t]{1.0\linewidth}
\centering
\includegraphics[width=2.6in]{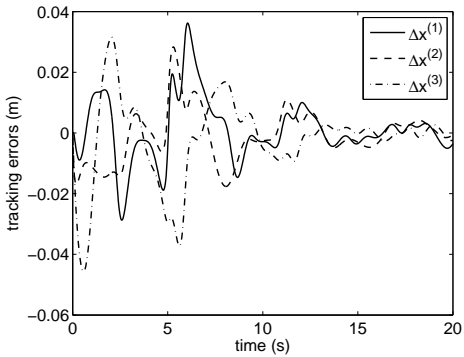}
\caption{Task-space position tracking errors (observer-based tracking control).}\label{fig:side:a}
\end{minipage}%
\end{figure}

\begin{figure}
\centering
\begin{minipage}[t]{1.0\linewidth}
\centering
\includegraphics[width=2.6in]{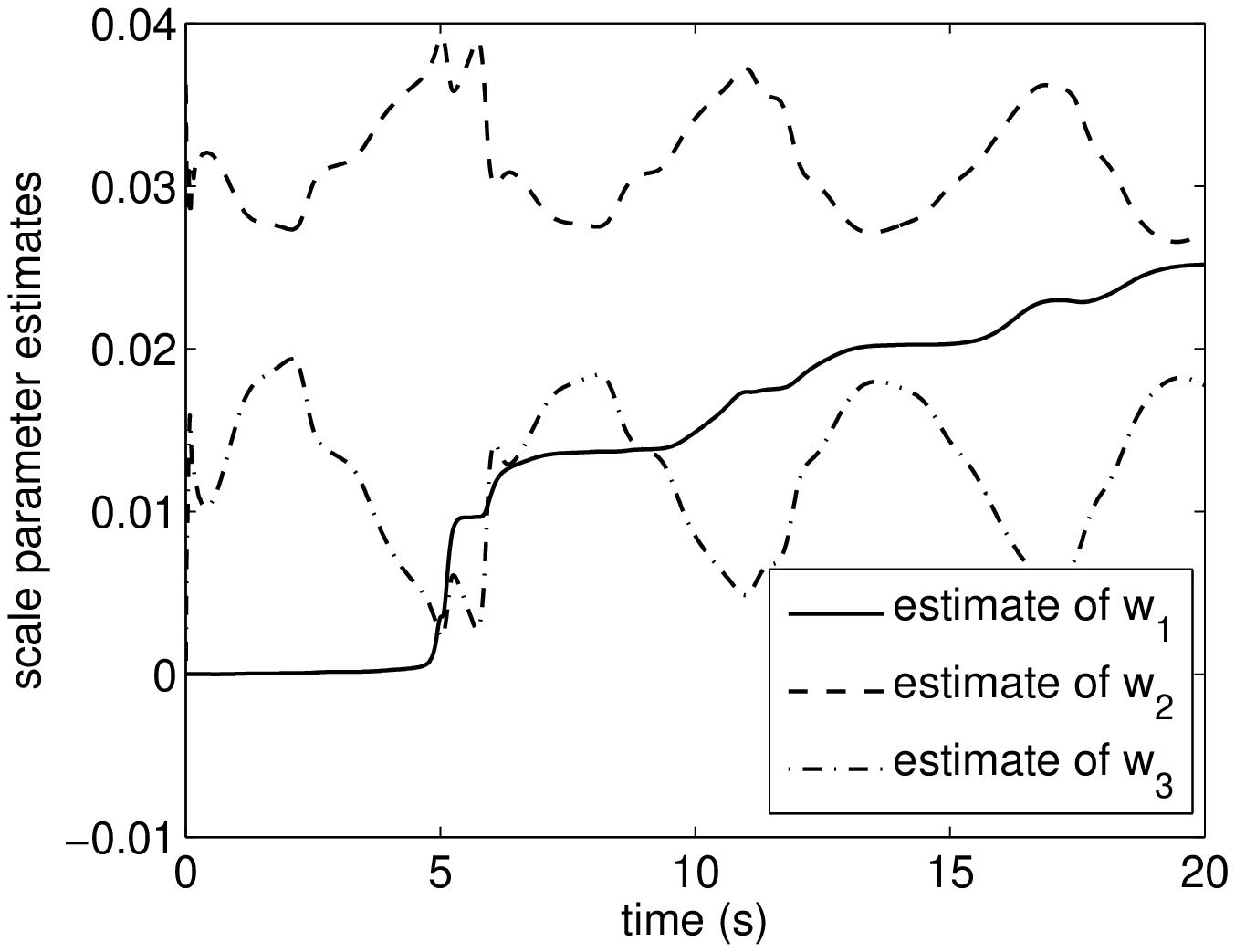}
\caption{Scale parameter estimates (observer-based tracking control).}\label{fig:side:a}
\end{minipage}%
\end{figure}

\begin{figure}
\centering
\begin{minipage}[t]{1.0\linewidth}
\centering
\includegraphics[width=2.6in]{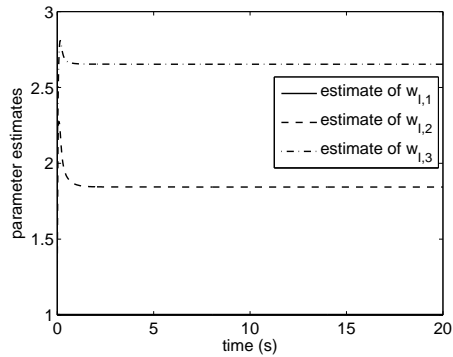}
\caption{Estimate of $w_I$ (observer-based tracking control).}\label{fig:side:a}
\end{minipage}%
\end{figure}

\subsection{Joint-Space Adaptive Control}

Consider first the outer loop version of the direct adaptive controller proposed by Slotine and Li with the controller parameters being chosen as $\bar\alpha=2.0$, $\Lambda=0.5 I_3$, $\Gamma_d=0.5 I_{15}$, and $\Lambda_I=100.0I_3$. The desired joint trajectory is set as $q_d=36[(1-\cos \pi t),\sin\pi t, \sin \pi t]^T$ deg. The initial parameter estimates are chosen as $\hat w(0)=0_3$, $\hat a_d(0)=0_{15}$, and
$\hat w_I(0)=\left[1.0,1.0,1.0\right]^T$. The joint position tracking errors and parameter estimates are shown in Fig. 12, Fig. 13, and Fig. 14.

\begin{figure}
\centering
\begin{minipage}[t]{1.0\linewidth}
\centering
\includegraphics[width=2.6in]{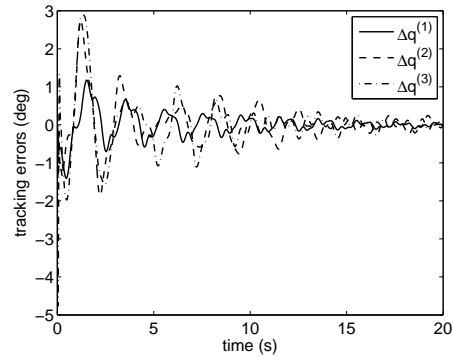}
\caption{Joint position tracking errors (direct adaptive controller).}\label{fig:side:a}
\end{minipage}%
\end{figure}

\begin{figure}
\centering
\begin{minipage}[t]{1.0\linewidth}
\centering
\includegraphics[width=2.6in]{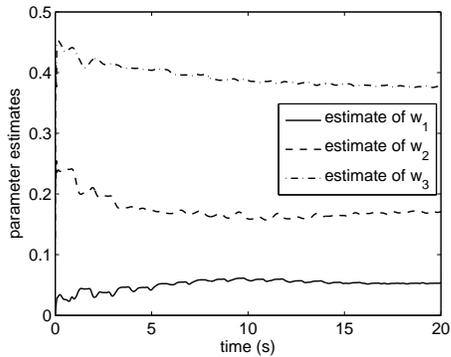}
\caption{Scale parameter estimates (direct adaptive controller).}\label{fig:side:a}
\end{minipage}%
\end{figure}

\begin{figure}
\centering
\begin{minipage}[t]{1.0\linewidth}
\centering
\includegraphics[width=2.6in]{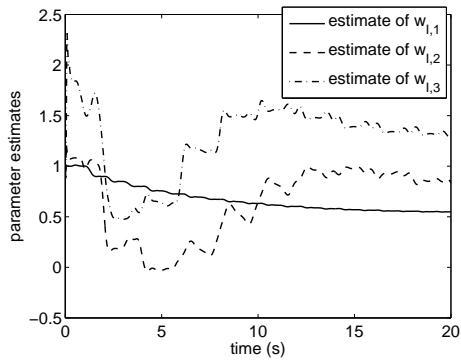}
\caption{Estimate of $w_I$ (direct adaptive controller).}\label{fig:side:a}
\end{minipage}%
\end{figure}

We next consider the case of using the composite adaptation in Sec. IV-B with the controller parameters $\gamma_0$ and $\lambda_f$
being chosen as $\gamma_0=0.3$ and $\lambda_f=1.0$ and the other controller parameters the same as those of the direct adaptive controller. The simulation results are shown in Fig. 15, Fig. 16, and Fig. 17, and in comparison with Fig. 12, Fig. 13, and Fig. 14, we see smoother tracking errors and parameter estimates as well as the improved convergence of the tracking errors.

\begin{figure}
\centering
\begin{minipage}[t]{1.0\linewidth}
\centering
\includegraphics[width=2.6in]{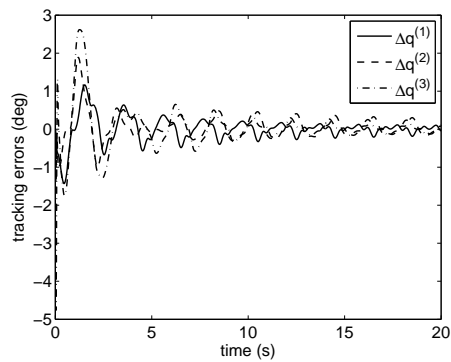}
\caption{Joint position tracking errors (composite adaptive controller).}\label{fig:side:a}
\end{minipage}%
\end{figure}

\begin{figure}
\centering
\begin{minipage}[t]{1.0\linewidth}
\centering
\includegraphics[width=2.6in]{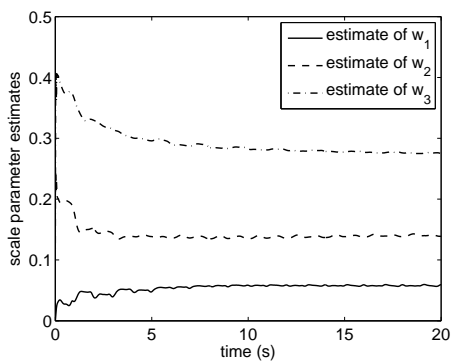}
\caption{Scale parameter estimates (composite adaptive controller).}\label{fig:side:a}
\end{minipage}%
\end{figure}

\begin{figure}
\centering
\begin{minipage}[t]{1.0\linewidth}
\centering
\includegraphics[width=2.6in]{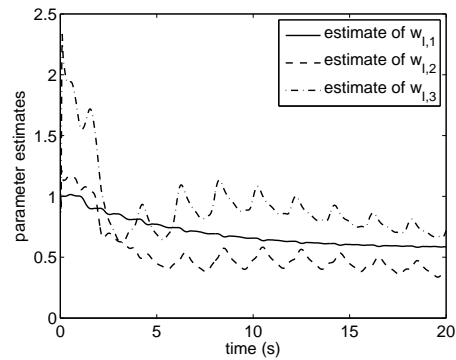}
\caption{Estimate of $w_I$ (composite adaptive controller).}\label{fig:side:a}
\end{minipage}%
\end{figure}

\subsection{Joint-Space Adaptive Control Considering Joint Flexibility}

The joint stiffness matrix $K_s$ is set as $K_s=10^6\times {\rm diag}[6.0,3.0,1.0]$ and the diagonal matrix $B_r$ is set as $B_r={\rm diag}[0.30, 0.20, $ $0.15]$. The controller parameters are chosen to be the same as the rigid manipulator case except that the adaptation gain $\Lambda_I$ is reduced to $\Lambda_I=60.0I_3$. The simulation results are shown in Fig. 18, and we see that the performance is comparable with the rigid manipulator case. But one may need to be cautious about the choice of the rotor inertias, and if the rotor inertias are too small [compared with the manipulator inertia matrix due to the link motion, i.e., $M_0(q)$ in (\ref{eq:72})], it is hard to choose a group of controller parameters that can stabilize the system. This is understandable as we recall the standard practice in terms of the design of the motor inertia, i.e., in the case that the motor inertia is strikingly smaller than the load inertia, the whole system would be quite difficult to stabilize and the use of advanced control algorithms does not help much. In particular, we perform a simulation for the case of reduced joint stiffness matrix, i.e., setting $K_s$ to be $K_s=10^4\times {\rm diag}[6.0,3.0,1.0]$ with the rotor inertias remaining unchanged. In addition, as the joint becomes more flexible, it is hard for the manipulator to track a fast time-varying trajectory, and therefore we slow down the evolution of the desired trajectory as $q_d=36[1-\cos (\pi t/3),\sin(\pi t/3), \sin (\pi t/3)]^T$ deg. The joint tracking errors are shown in Fig. 19. But as we reduce the rotor inertias to, e.g., 50\% of the original, it is very difficult to stabilize the system even with the joint stiffness being unchanged (relatively high). To illustrate the reason behind this phenomenon, we calculate the link inertia matrix $M_0(q)$ at $q=q(0)=0_3$ and its value is $$M_0(q)=\begin{bmatrix}18.9058 &        0         &0\\ 0 &  18.9290  &  9.4327\\ 0 &   9.4327  &  5.1205\end{bmatrix}.$$ The three eigenvalues of $M_0(q)$ are $ 0.3352$,  $ 18.9058$,  and $23.7143$. This means that the maximum load/rotor inertia ratio reaches $23.7143/0.3\approx79$ (happening at the second joint) and such a ratio makes it challenging to stabilize the system. One solution to this problem, as suggested by the standard results in the design of the load/motor ratio, is to increase the joint stiffness, and in other words, larger stiffness allows the specification of larger load/rotor inertia ratio. Fig. 20 shows the joint tracking errors as the stiffness is increased to $K_s=10^8\times {\rm diag}[6.0,3.0,1.0]$ with the desired joint position being still the same as the rigid manipulator case. However, in the simulation, we have to decrease the integration step size so that the stability of the numerical integration can be ensured since the degree of stiffness of the system dynamics is increased. In this particular case (i.e., increased joint stiffness and decreased rotor inertia), the step size is decreased from $0.5$ ms to $0.05$ ms.

\begin{figure}
\centering
\begin{minipage}[t]{1.0\linewidth}
\centering
\includegraphics[width=2.6in]{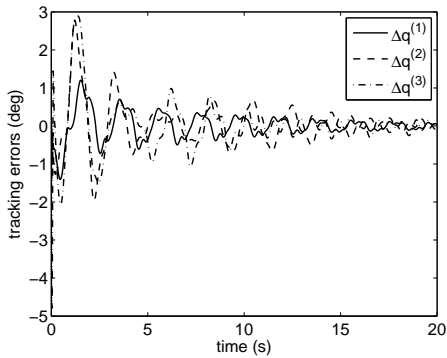}
\caption{Joint position tracking errors (flexible joint case).}\label{fig:side:a}
\end{minipage}%
\end{figure}

\begin{figure}
\centering
\begin{minipage}[t]{1.0\linewidth}
\centering
\includegraphics[width=2.6in]{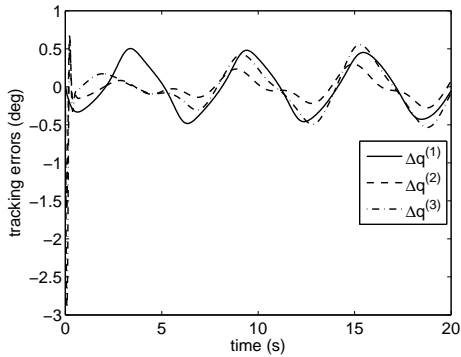}
\caption{Joint position tracking errors (reduced joint stiffness).}\label{fig:side:a}
\end{minipage}%
\end{figure}

\begin{figure}
\centering
\begin{minipage}[t]{1.0\linewidth}
\centering
\includegraphics[width=2.6in]{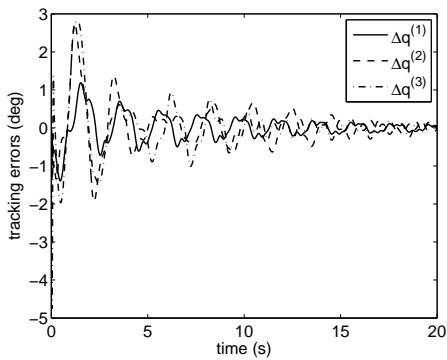}
\caption{Joint position tracking errors (increased joint stiffness and decreased rotor inertias).}\label{fig:side:a}
\end{minipage}%
\end{figure}

\subsection{Joint-Space Adaptive Control With an Inner PID Position Controller}

The gains of the inner PID position controller is set as $K_D=30.0I_3$, $K_P=15.0 I_3$, and $K_I=10.0 I_3$. The controller parameters for the joint-space adaptive outer loop controller (which can be developed similarly to the one in Sec. III-D) are chosen as $\bar\alpha=1.5$, $K_c=0.8$,
$\Gamma_d=0.5I_{15}$, $\Lambda=0.5 I_3$, $\Lambda_P=100.0 I_3$, and $\Lambda_I=50.0 I_3$. The initial parameter estimates are chosen as
$\hat a_d(0)=0_{15}$, $\hat w(0)=0_3$, $\hat w_P(0)=[1.0,1.0,1.0]^T$, and $\hat w_I(0)=[1.0,1.0,1.0]^T$. The desired trajectory is set to
be the same as
the case of an inner PI velocity controller. The joint position tracking errors are shown in Fig. 21, which is comparable with
the case of an inner PI velocity controller (Fig. 12). Interestingly, no unstable phenomenon is observed even with very fast adaptation to ${\mathcal K}_P$ and ${\mathcal K}_I$ although it is currently still challenging to rigorously
ensure the uniform exponential stability of (\ref{eq:a1}), and the evolution of $q_c-q_r$ is plotted in Fig. 22. The quantity $q_c-q_r$ (which is
apparently bounded based on the data shown in Fig. 22) characterizes the injected dynamic compensation of the proposed controller.

\begin{figure}
\centering
\begin{minipage}[t]{1.0\linewidth}
\centering
\includegraphics[width=2.6in]{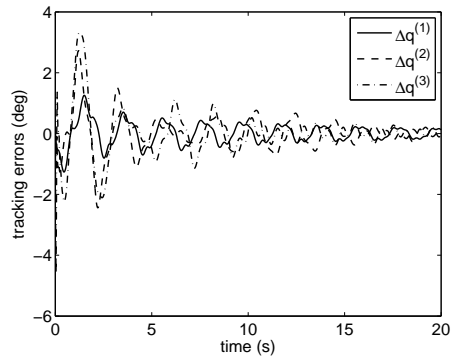}
\caption{Joint position tracking errors (adaptive outer loop controller with an inner PID position controller).}\label{fig:side:a}
\end{minipage}%
\end{figure}

\begin{figure}
\centering
\begin{minipage}[t]{1.0\linewidth}
\centering
\includegraphics[width=2.6in]{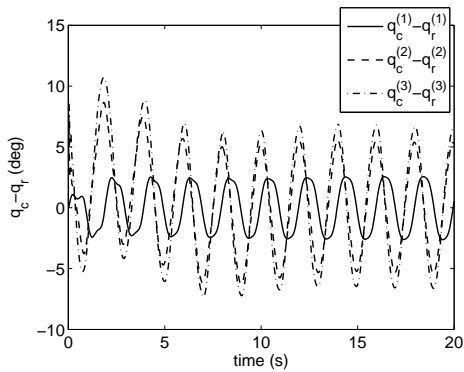}
\caption{$q_c-q_r$ (adaptive outer loop controller with an inner PID position controller).}\label{fig:side:a}
\end{minipage}%
\end{figure}

\subsection{Further Remarks}

One key issue in the above simulations is the choice of controller parameters including those of the low-level PI controller. Different from most theoretical results in the literature that design control laws at the torque level, the system here actually has two loops with strikingly different updating frequency, namely, the low-rate outer loop and the high-rate inner joint servoing loop. Due to this structure, to guarantee the robustness and performance of the whole system, high gains are specified in the high-rate joint servoing loop while low gains are specified in the low-rate outer loop. We naturally produce a system that consists of two loops with two time-scales and interestingly, it is the system constraint that gives rise to the two-time-scale behavior of the closed-loop system that we cannot modify.

Another issue often involved in practice is the computational efficiency, especially in the case that the number of the DOFs of the manipulator is very large. The typical solution to this problem is the recursive implementation of the adaptive controllers---see, e.g., \cite{Niemeyer1991_IJRR,Wang2010_CCC,Wang2013_ASME}. The recursive direct adaptive controller as detailed in \cite{Niemeyer1991_IJRR,Wang2010_CCC} has the complexity $O(n)$ and the complexity of the recursive composite adaptive controller in \cite{Wang2013_ASME} is $O(n^2)$, where $n$ denotes the number of the DOFs of the manipulator. We may also note that all the complicated computations of the nonlinear and coupling terms take place in the low-rate outer loop with a powerful computer, and thus the complexity up to $O(n)$ or $O(n^2)$ is expected to be acceptable.

\section{Experimental Results}

To further show the practical performance of the proposed adaptive outer loop controllers, we perform the experimental study using the UR10 of Universal Robots at Nanyang Technological University (see Fig. 23). We fix the position of upper three joints of the UR10 robotic system to be $[0,-1.57,0]^T$ rad, and only the motion of the first three DOFs is considered, and the low-level inner loop controller is considered to take the PID position control action. Either the position or velocity command for the UR10 robot can be designed and it is not allowed to deign the position and velocity commands simultaneously. The sampling period of the outer loop is around (not exactly) 0.2 s.
\begin{figure}
\centering
\begin{minipage}[t]{1.0\linewidth}
\centering
\includegraphics[width=2.6in]{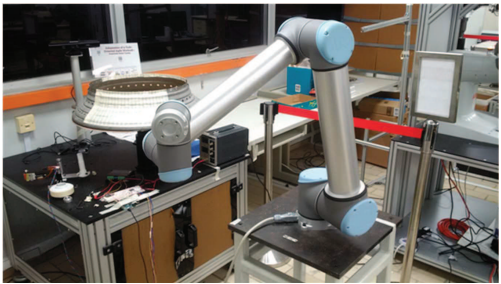}
\caption{Experimental setup with UR10.}\label{fig:side:a}
\end{minipage}%
\end{figure}

We here only take into consideration the Cartesian-space tracking control problem with unknown system dynamics and unknown low-level PID controller parameters while the kinematics of the system is accurately known. That is, we use the joint reference velocity defined by (\ref{eq:39}) yet with the estimated Jacobian matrix being replaced by the actual Jacobian matrix and with $x_o$ being replaced by $x$. The desired trajectory in the Cartesian space is specified as
\be
x_d(t)=\begin{bmatrix}-0.45+0.1\sin(t)\\0.45+0.1\cos(t)\\0.5\end{bmatrix}.
\ee
The controller parameters are chosen as $\gamma=0.5$, $K_c=0.8$, $\Gamma_d=1.0 I_{16}$, $\Lambda=0.01 I_3$, $\Lambda_P=0.02I_3$, and $\Lambda_I=0.02 I_3$. The initial values of the dynamic parameter estimates are set to be zero, i.e., $\hat a_d(0)=0_{16}$. Note that here
we expand $a_d$ in (\ref{eq:5}) to include the parameters associated with the Coulomb friction which can be written as \cite{Slotine1991_Book}
\be
f_c=D{\rm sgn}(\dot q)
\ee
where ${\rm sgn}(\dot q)=\left[{\rm sgn}(\dot q_1),{\rm sgn}(\dot q_2), {\rm sgn}(\dot q_3)\right]^T$ and $D$ is a $3\times 3$ diagonal positive definite matrix with its diagonal entries unknown. Correspondingly, the dynamic regressor matrix used in the experiment is the combination of $Y_d(q,\dot q,\dot q_r,\ddot q_r)$ and ${\rm diag}[{\rm sgn}(\dot q_{r})]$. The initial values of $w$, $w_P$, and $w_I$ are chosen as $\hat w(0)=0_3$, $\hat w_P(0)=[1.8,1.8,1.8]^T$, and $\hat w_I(0)=[1.2,1.2,1.2]^T$, respectively. The position tracking errors as using the proposed controller are shown in Fig. 24. For comparison, the position tracking errors as using $q_r$ as the joint position command (i.e., the conventional kinematic controller) are shown in Fig. 25. The tracking accuracy is apparently improved by using the proposed dynamic modularity approach.

\begin{figure}
\centering
\begin{minipage}[t]{1.0\linewidth}
\centering
\includegraphics[width=2.6in]{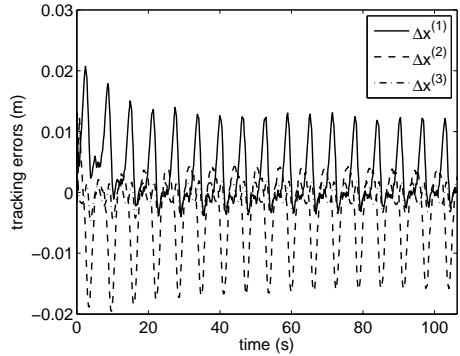}
\caption{Position tracking errors (using $q_c$ as the joint position command).}\label{fig:side:a}
\end{minipage}%
\end{figure}

\begin{figure}
\centering
\begin{minipage}[t]{1.0\linewidth}
\centering
\includegraphics[width=2.6in]{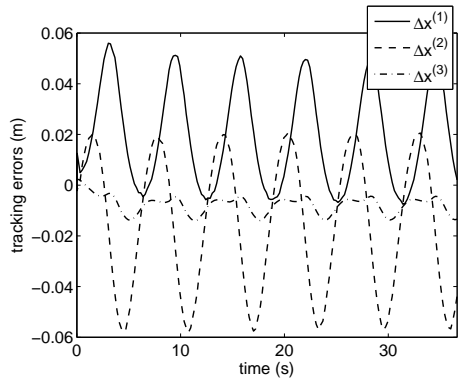}
\caption{Position tracking errors (using $q_r$ as the joint position command, i.e., the conventional kinematic controller).}\label{fig:side:a}
\end{minipage}%
\end{figure}

\section{Conclusion}

In this paper, we have proposed a dynamic modularity approach to adaptive control of robotic systems with an inner/outer loop structure, and both the task-space and joint-space control are taken into consideration under this framework. The proposed adaptive outer loop controllers take into full account the system dynamic effects while most existing kinematic controllers rely on the ad hoc assumption of fast enough joint servoing loop or the modification of the low-level joint servoing controller to be much more complicated one. From an application perspective, most existing results cannot ensure the stability of the system or convergence of the tracking/regulation error as applied to robotic systems with an inner/outer loop structure (e.g., most commercial/industrial robotic systems) while the proposed adaptive outer loop schemes can guarantee the stability and convergence of the system without the need to modify the low-level joint servoing loop. The goal of the study here is to yield a module robot control system where the adaptive outer loop is user-defined and the inner loop is factory-defined and embedded.

It might be worth discussing the roles of feedback separation in the proposed controllers. Feedback separation is a design objective introduced in the context of Cartesian-space control and visual servoing control of robots with uncertain kinematics (see, e.g., \cite{Wang2016_TAC,Wang2015_TACsubmitted}). Feedback separation is initially for generating simple adaptive kinematic controllers for industrial/commertial robots, and on the other hand it can potentially reduce the activity of the dynamic compensation action. For instance, the filter-based regulation algorithm in Sec. III-A actually does not achieve feedback separation and thus the scaled dynamic compensation action given in (\ref{eq:11}) involves an additional term $-\alpha \hat J^T(q)\Delta x$ (i.e., a stronger compensation action is required). In contrast, both the observer-based regulation and tracking controllers achieve the feedback separation, benefiting from which, the scaled dynamic compensation action defined in (\ref{eq:23}) no longer involves additional terms (of course, the controller structure becomes more complex since an observer is introduced). From a control viewpoint, this leads us to reconsider the issue of the cancellation of indefinite terms in the standard backstepping-based control. The cancellation of indefinite terms can lead to a good form of the derivative of the Lyapunov function but often gives rise to potentially decreased robustness and strong coupling between different control loops. The realization of feedback separation may help avoid the unfavorable cancellations of indefinite terms.

Furthermore, the proposed approach may possibly be applicable to other classes of (commercial) mechanical systems (e.g., space robots, mobile robots, or aerial vehicles) that have a hidden torque/force control loop yet admit the design of the velocity (or position) command. 


%

%



\section*{Acknowledgment}

The authors would like to thank Dr. Tiantian Jiang, Dr. Yong Hu, and Dr. Yong Wang for the helpful discussions and comments on stability issues concerning the linear time-varying systems.





\bibliographystyle{IEEEtran}
\bibliography{..//Reference_list_Wang}

%
%
%

%








\end{document}